\documentclass[12pt,a4paper,dvips]{article}

\usepackage{textcomp}
\usepackage[latin1]{inputenc}
\usepackage{pstricks,pst-node,pst-text}

\usepackage{amssymb}
\usepackage{amsthm}
\usepackage{amsmath}
\usepackage{amsfonts,color,graphicx,cite}
\usepackage{caption}
\usepackage[labelfont=bf,textfont={sl,bf}]{subfig}
\usepackage{verbatim}
\usepackage{dcolumn}

\newcommand{\RE}{\operatorname{Re}}

\newcommand{\cO}{{\cal O}}

\newcommand{\dd}{{\textrm{d}}}
\newcommand{\hsigma}{{\hat{\sigma}}}

\newcommand{\be}{\begin{equation}}
\newcommand{\ee}{\end{equation}}
\newcommand{\bea}{\begin{eqnarray}}
\newcommand{\eea}{\end{eqnarray}}
\newcommand{\ba}{\begin{array}}
\newcommand{\ea}{\end{array}}
\newcommand{\ben}{\begin{enumerate}}
\newcommand{\een}{\end{enumerate}}
\newcommand{\beal}{\begin{aligned}}
\newcommand{\eeal}{\end{aligned}}
\newcommand{\ban}{\begin{align}}
\newcommand{\ean}{\end{align}}
\newcommand{\bitem}{\begin{itemize}}
\newcommand{\eitem}{\end{itemize}}

\newcommand{\bc}{\begin{center}}
\newcommand{\ec}{\end{center}}

\newcommand{\bpmatrix}{\begin{pmatrix}}
\newcommand{\epmatrix}{\end{pmatrix}}
\newcommand{\bbmatrix}{\begin{bmatrix}}
\newcommand{\ebmatrix}{\end{bmatrix}}

\newcommand{\crn}{\nonumber \\}
\newcommand{\fr}{\frac}
\newcommand{\de}{\delta}
\newcommand{\De}{\Delta}
\newcommand{\al}{\alpha}
\newcommand{\la}{\lambda}

\newcommand{\ga}{\gamma}

\newcommand{\eps}{\epsilon}

\newcommand{\si}{\sigma}

\newcommand{\bbWHp}{b\bar{b}\to W^{-}H^{+}}

\newcommand{\ppHtbm}{pp\to H^{-}t\bar{b}}

\newcommand{\ggHtbm}{gg\to H^{-}t\bar{b}}

\newcommand{\ggHtbmp}{gg\to H^{\mp}tb}

\newcommand{\qqHtbm}{q\bar q\to H^{-}t\bar{b}}

\newcommand{\bbHtbm}{b\bar b\to H^{-}t\bar{b}}

\newcommand{\gyHtbm}{g\ga\to H^{-}t\bar{b}}
\newcommand{\ggHtby}{gg\to H^{-}t\bar{b}\ga}

\newcommand{\gbHmpt}{gb\to H^{\mp}t}

\def\slashc{c\kern -.400em {/}}
\def\slashp{p\kern -.400em {/}}
\def\slashq{q\kern -.450em {/}}
\def\slashL{L\kern -.450em {/}}
\def\slashcl{\cl\kern -.600em {/}}
\def\slashr{r\kern -.450em {/}}
\def\slashk{k\kern -.500em {/}}
\def\slasheta{\eta\kern -.500em {/}}
\def\slashep{\epsilon\kern -.450em {/}}
\def\slashpbar{\bar{p}\kern -.450em {/}}
\def\slashD{D\kern -.650em {/}}
\def\slashepi{\epsilon_i\kern -.720em {/}}
\def\slashpi{p_i\kern -.600em {/}}

\newcommand{\hs}{\hspace*{3mm}}

\newcommand{\ie}{{\it i.e.\;}}
\newcommand{\eg}{{\it e.g.\;}}

\newcommand{\gev}{~\text{GeV}}

\newcommand{\tev}{~\text{TeV}}

\newcommand{\DRb}{\overline{\text{DR}}}

\newcommand{\MSb}{\overline{\text{MS}}}

\newcommand{\tbeta}{{\tan{\beta}}}
\newcommand{\MHpm}{{M_{H^\pm}}}

\newcommand{\eq}[1]{Eq.~(\ref{#1})}
\newcommand{\fig}[1]{Fig.~\ref{#1}}
\newcommand{\tab}[1]{Table~\ref{#1}}
\newcommand{\sect}[1]{Section~\ref{#1}}

\newcommand{\mbmb}{\overline{m}_b(\overline{m}_b)}

\newcommand{\mbDRb}{m_b^{\DRb}}

\newcommand{\ltff}{{\texttt{LoopTools/FF}}}
\newcommand{\lis}{{\texttt{LoopInts}}}

\newcommand{\fcv}{{\texttt{FormCalc-6.0}}}

\newcommand{\fav}{{\texttt{FeynArts-3.4}}}

\newcommand{\bases}{{\texttt{BASES}}}

\newcommand{\vegas}{{\texttt{VEGAS}}}

\def\slashepi{\epsilon_i\kern -.720em {/}}
\def\slashpi{p_i\kern -.600em {/}}
\def\slashp{p\kern -.550em {/}}

\begin{document}
\begin{titlepage}
\vspace*{-1cm}
\rightline{KA-TP-37-2012}
\rightline{MPP-2012-137}
\rightline{SFB/CPP-12-75}

\vspace*{2cm}
\begin{center}

{\Large{\bf Electroweak corrections to {\boldmath$\ggHtbm$} at the LHC}}

\vspace{.5cm}

DAO Thi Nhung$^{a}$, Wolfgang HOLLIK$^b$ and LE Duc Ninh$^{a}$

\vspace{4mm}
{\it $^a$Institut f\"ur Theoretische Physik, Karlsruher Institut f\"ur  Technologie, \\
D-76128 Karlsruhe, Germany}

{\it $^b$Max-Planck-Institut f\"ur Physik (Werner-Heisenberg-Institut), \\
D-80805 M\"unchen, Germany}

\vspace{10mm}
\abstract{ 
The dominant contribution to $H^- t\bar{b}$ production at the LHC is the  
gluon-gluon fusion parton subprocess.
We perform for the case of the complex MSSM
a complete calculation of the NLO electroweak contributions to this channel. 
The other small contributions with quarks or photon in the initial state are 
calculated at tree level. The results are improved by using the effective bottom-Higgs 
couplings to resum the leading radiative corrections. We find that,
beyond these  leading corrections,  the NLO electroweak contributions can be still 
be significant. 
The effect of the complex phases of the soft-breaking parameters is found to 
be sizeable.
}

\end{center}


\normalsize

\end{titlepage}


\section{Introduction}
Charged Higgs boson production 
in association with a top quark is the dominant mechanism in charged-Higgs 
searches at the LHC. The leading order (LO) tree-level diagrams involve a gluon and 
a bottom quark in the initial state. 
The calculation of the cross section 
can be performed in two ways, by using the four- or the five-flavor schemes. 
In the 4-flavor scheme (4FS), the bottom density is zero and the leading
contribution is $\ggHtbmp$ 
whose total cross section contains large logarithm $\sim\ln\mu_F/m_b$,
where the factorization scale $\mu_F$ is of the order of the charged Higgs mass. 
This correction arises from the splitting 
of a gluon into a collinear $b\bar b$ pair. 
In the 5-flavor scheme (5FS) the bottom density is non-zero and the leading contribution is $\gbHmpt$. 
The large collinear corrections are resummed to all orders and are included in the bottom distribution 
functions. The two schemes should give the same result for the total cross section if the calculations 
are done to a sufficiently high order in perturbation theory. 
A comparison at next-to-leading order (NLO) has been done in \cite{Dittmaier:2009np}. The
results of the two schemes are consistent within the scale
uncertainties, with the central predictions in the 5FS 
being larger than those of the 4FS \cite{Dittmaier:2009np}. 

From an experimental point of view, the two final states $H^\mp t$ and $H^\mp tb$ 
can be separated by requiring $b$ tagging. For a heavy charged Higgs boson 
($M_H^\pm > m_t$) decaying into $tb$, the signal contains $3b$s for the 
former and $4b$s for the latter. In general, the addition of a bottom quark to 
the final state reduces the signal rate, but the background is also lowered. 
The study in \cite{Miller:1999bm} (see also \cite{Roy:2005yu} and references therein) shows that a good signal-to-background ratio 
can be achieved by imposing $4$ $b$-tags and suitable cuts if $M_{H^\pm}$ is significantly 
larger than $m_t$. This study, however, is based on LO predictions and the large $\tan\beta$ (the 
ratio of the two vacuum expectation values of the two Higgs doublets) enhanced corrections to the bottom-Higgs couplings are not taken into account. 
Those large corrections, which 
can be resummed and easily included to the LO results by using the effective bottom-Higgs couplings, 
can significantly change the signal cross section, in particular for larger values of $\tan\beta$. 
It is therefore important to know the quality of this approximation
and to have some idea about the remaining higher-order uncertainty. 
A comparison with the full NLO results is needed. 

In the Minimal Supersymmetric Standard Model (MSSM), the 
NLO corrections to charged Higgs production in association with heavy
quarks at the LHC have been studied to some extent. 
For the $H^\mp t$ production, both the QCD and the electroweak (EW) NLO corrections 
have been calculated \cite{Zhu:2001nt, Gao:2002is, Plehn:2002vy,Berger:2003sm, Beccaria:2009my}, 
and some higher-order QCD corrections in \cite{Kidonakis:2005hc, Kidonakis:2010ux}. 
For the exclusive $H^\mp tb$ production, the QCD corrections have been
considered in \cite{Peng:2006wv, Dittmaier:2009np}, and the supersymmetric (SUSY) QCD corrections 
for $e^+e^-$ collider in \cite{Kniehl:2010ea}. 
The EW corrections are missing. 
All those studies assume that the soft-breaking parameters are real. 

The purpose of this paper is to provide \footnote{The computer code can be obtained from the authors upon request.} 
and study the EW corrections to the exclusive $H^{-}t\bar{b}$ 
production at the LHC for heavy $H^\pm$ (with $M_H^\pm > m_t$). 
The tagged bottom quark is required to satisfy the kinematic constraint: 
\be p_{T,b} > 20\gev,\quad |\eta_b|<2.5, \label{eq_cuts} \ee 
where $p_{T,b}$ is the transverse momentum and $\eta_b$ is the pseudorapidity. 
The cross section after cuts is still considerable. Our study is done in the MSSM with complex parameters (complex MSSM, or cMSSM). 
The impact of the important phases on the cross section will be quantified. It turns out that this effect is not small. 

The paper is organized as follows. \sect{sec:LO_ggHtb} is devoted to the tree-level study, 
including the issue of the effective bottom-Higgs couplings. The 
calculation of the NLO EW corrections to the process $\ggHtbm$ is done in \sect{sec:NLO_ggHtb}. 
Numerical results are presented 
in \sect{sect-results} and conclusions in~\sect{sect-conclusions}.
 
\section{Leading order consideration}
\label{sec:LO_ggHtb}
At tree level, the $gg$ contributions of order $\cO(\al_s^2\al)$ are dominant.  Other
 contributions 
of the same order arising from $q\bar{q}$ ($q$ is a light quark) 
annihilations are much smaller, since they involve
only the $s$ channel diagrams which are suppressed at high energy 
and the quark density is smaller than the gluon one at the LHC. 
We will, however, include those contributions at tree level. 
It is noted that 
the $q\bar{q}$ annihilations give also $\cO(\al^3)$ contributions
coming from the tree-level EW Feynman diagrams. 
These small channels are neglected in our calculation. 

\begin{figure}[h]
  \centering
 \subfloat[]{ \includegraphics[width=0.8\textwidth,height=0.30\textwidth]
{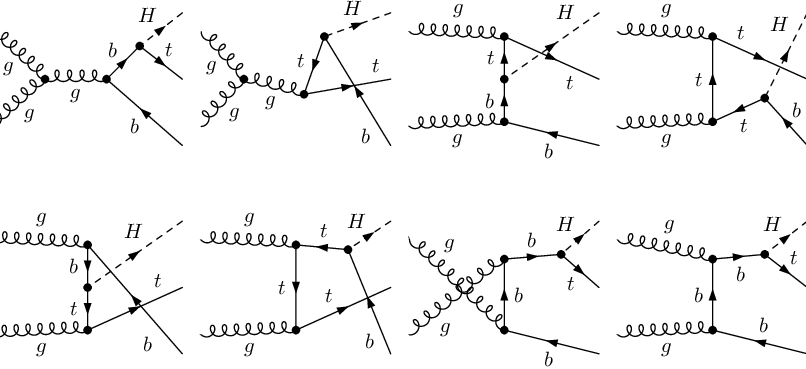}}\\
  \subfloat[ ]{  \includegraphics[width=0.4\textwidth,height=0.14\textwidth]
{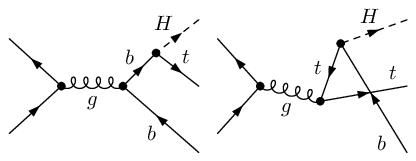}}\\
  \subfloat[ ]{  \includegraphics[width=0.8\textwidth,height=0.14\textwidth]
{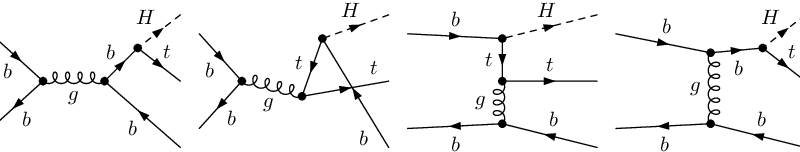}}
  \caption{The QCD tree-level diagrams: (a) for  the $\ggHtbm$
    subprocess, (b) for  the $\qqHtbm$ subprocesses ($q=u,c,d,s$) 
    and (c) for the $\bbHtbm$ subprocess.}
  \label{ppHtb_QCD_tree_diagrams}
\end{figure}

We assume the 5FS with $b$ tagging (see the discussion below). 
The three classes of subprocesses of order $\cO(\al_s^2\al)$ are
\begin{align}
g+g &\to H^-+ t+\bar{b},\label{eq:ggHtb}\\ 
q+\bar{q} &\to H^-+ t+\bar{b},\label{eq:qqHtb}\\
b+\bar{b}&\to H^-+ t+\bar{b}\label{eq:bbHtb},
\end{align} 
where $q=u,c,d,s$. The first two channels have been calculated 
in \cite{DiazCruz:1992gg, Borzumati:1999th, Miller:1999bm,
  Peng:2006wv, Dittmaier:2009np}. 
The corresponding Feynman diagrams of those subprocesses are shown in \fig{ppHtb_QCD_tree_diagrams}. 
The last process is expected to be small and will be shown to be numerically irrelevant. 
It should be noted that the $b\bar{b}$ annihilation containing the collinear splitting 
$b\to bg^*$ is suppressed by the $p_{T,b}$ cut. 

\begin{figure}[h]
  \centering
   \includegraphics[width=01\textwidth,height=0.150\textwidth]
{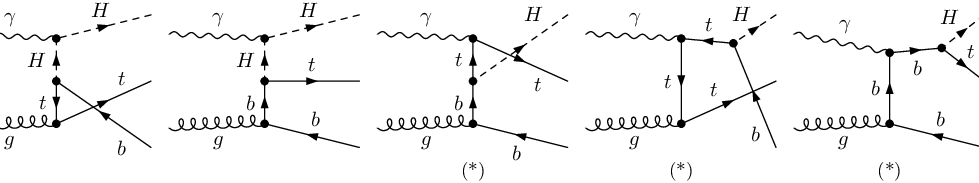}
  \caption{The tree-level diagrams for the
$\gyHtbm$ subprocess. The star means that the graphs with  the two incoming particles 
interchanged also contribute.}
 \label{gyHtb_EW_tree_diagrams}
\end{figure}

There exists also a contribution of order $\cO(\al_s\al^2)$ arising from the photon-induced process,
\be g+\ga \to H^-+ t+\bar{b}\label{eq:gyHtb}, \ee
according to the  Feynman diagrams  depicted in \fig{gyHtb_EW_tree_diagrams}. 
Compared to the  $gg$ fusion, 
a new EW splitting $\gamma \to H^+ H^-$ appears. This splitting leads to 
contributions increasing with decreasing $M_H^\pm$. 
Although the $g\gamma$ cross section is larger than the one from $b\bar{b}$ annihilation,
it  turns out to be negligible as well, 
as we will show in our numerical analysis. The small 
$\gamma\gamma$ fusion contribution of $\cO(\al^3)$ is neglected. 

We have a few comments on the 5FS assumption.
In this paper, we are primarily concerned with the NLO EW corrections
to the process $gg \rightarrow H^- t\bar{b}$;
the issue of choosing the 4FS or the 5FS is numerically not important
in this context. Basically, also the EW contributions are
affected by this choice, since taking into account photon splitting
into $b\bar{b}$ pairs in the evolution defines the 5FS scheme also
in the context of QED. This point, however, is not relevant at the
EW NLO level because the contributions arising from
initial-state photons are small and the differences of the two schemes
in the evolution are small, too.
Using the 5FS here means in practice that we include the subprocess
with initial-state bottom quarks at tree level
(which is negligible as above said) 
and use the 5FS parton distribution functions (PDF) from the MRST2004qed set~\cite{Martin:2004dh} 
which includes the EW effects and the photon density in the proton. 
This is, however, not an ideal choice for calculating 
the exclusive $H^-t\bar{b}$ production rate at the LHC. 
The use of the 5FS PDFs with large factorization scale $\mu_F\approx M_{H^\pm}$ 
implies that our calculation includes also the contributions with 
more than one $b$ quarks in the final state. Since these higher-order 
$b$ corrections enter in the same factorization manner in both the LO 
and the NLO results, they are expected to have a minor impact on the 
relative EW corrections. To get the best theoretical prediction, one 
has to include also the QCD corrections and this should be done 
in the framework of the 4FS in order to have a clean 
exclusive $H^-t\bar{b}$ final state, as discussed in \cite{Dittmaier:2009np}. 

All tree-level diagrams involve the Yukawa couplings of the charged Higgs bosons to the
top and bottom quarks, which read as follows,
\bea
\lambda_{b\bar{t}H^+}&=&\fr{ie}{\sqrt{2}s_WM_W}\left(\fr{m_t}{\tbeta}P_L + m_b\tbeta P_R\right),\crn
\lambda_{t\bar{b}H^-}&=&\fr{ie}{\sqrt{2}s_WM_W}\left(m_b\tbeta P_L + \fr{m_t}{\tbeta} P_R\right),
\label{b_H_couplings_tree}
\eea
where $P_{L,R}=(1\mp \gamma_5)/2$, $s_W =\sin\theta_W$. 
It is known that these couplings can get large Standard Model (SM) QCD, SUSY-QCD and 
EW corrections. The SM-QCD corrections are absorbed by the replacement $m_b\to \mbDRb(\mu_R)$ with 
$\mu_R$ being the renormalization scale, \ie the running quark mass is used. The universal SUSY-QCD 
and EW corrections are resummed via the quantity $\Delta_b$. The exact definition of $\mbDRb(\mu_R)$ 
and $\Delta_b$ are given in \cite{Dao:2010nu}. We just want to emphasize here that 
the quantity $\Delta_b$ is proportional to $\tan\beta$ and depends 
on the mass of the SUSY particles. Including these corrections, the effective bottom-top-Higgs couplings read 
\cite{Carena:1999py, Dao:2010nu, Dao:2012phd}: 
\bea
\bar\lambda_{b\bar{t}H^+}&=&\fr{ie}{\sqrt{2}s_WM_W}\left(\fr{m_t}{\tbeta}P_L + \mbDRb\tbeta \Delta_b^{3*} P_R\right),\crn
\bar\lambda_{t\bar{b}H^-}&=&\fr{ie}{\sqrt{2}s_WM_W}\left(\mbDRb\tbeta \Delta_b^{3} P_L + \fr{m_t}{\tbeta} P_R\right),
\label{b_H_couplings_loop}
\eea  
where
\bea
\Delta_b^3 &=& \fr{1-\Delta_b/(\tbeta)^2}{1+\Delta_b}.
\eea 
The top-quark mass is considered as the pole mass which is an input parameter in our calculation. 
In the explicit one-loop calculations, we have to subtract the EW part  
of the $\Delta_b$ correction which has already been included in the tree-level contribution to avoid double
counting. This can formally
be done by adding the following counterterms 
\bea
\delta m_b^{H^+}&=&\mbDRb\left[1+\fr{1}{(\tbeta)^2}\right](\Delta m_b^{\text{SEW}})^* P_R,\crn  
\delta m_b^{H^-}&=&\mbDRb\left[1+\fr{1}{(\tbeta)^2}\right]\Delta m_b^{\text{SEW}} P_L
\label{dMB_subtraction}
\eea 
to $\de m_b$ in the corresponding bottom-Higgs-coupling counterterms, 
as listed in  Appendix~B of \cite{Dao:2010nu}. The definition of 
$\Delta m_b^{\text{SEW}}$ is also given in \cite{Dao:2010nu}. 
 
To quantify the 
$\Delta_b$ effect we define the improved Born approximation (IBA) where the effective 
couplings in \eq{b_H_couplings_loop} are used. 
The LO cross section is computed with the tree-level couplings
in \eq{b_H_couplings_tree} with $m_b=m_b^{\DRb}(\mu_R)$. 

At the end, from the various partonic cross sections, either at LO or IBA,
$\hsigma^{ij}_{\text{LO/IBA}}$,
we obtain the corresponding LO and IBA hadronic cross
sections in the following way,
\bea
\sigma^{pp}_{\text{LO/IBA}} &=&
\sum_{i,j} \fr{1}{1+\delta_{ij}}
\int \dd x_1\dd x_2\;  [\, F_i^{p}(x_1, \mu_F)F_j^{p}(x_2, \mu_F) \;
\hsigma^{ij}_{\text{LO/IBA}}(\alpha_s^2\alpha,\alpha_s\alpha^2,\mu_R)\crn
& &  \qquad \qquad  \qquad \qquad \qquad   
       +\;   (1\leftrightarrow 2) \, ] \, ,
\eea
where $(i,j) =$ $(q,\bar{q})$, $(b,\bar{b})$, $(g,g)$, $(g,\gamma)$; 
$F_{i}^p(x,\mu_F)$ denotes the distribution function 
of parton $i$ at momentum fraction $x$ and factorization scale $\mu_F$.

\section{NLO electroweak contributions to $\ggHtbm$}
\label{sec:NLO_ggHtb}
\begin{figure}[]
\begin{center}
\includegraphics[width=0.8\textwidth]
{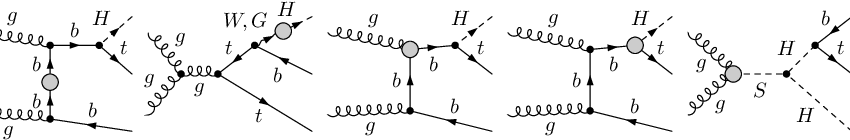}
\caption{Representative one-loop self-energy and vertex diagrams. 
The shaded regions are the one-particle irreducible two- and
three-point vertices
including the counterterms.
$G$ denotes the $W$ Goldstone bosons.} 
\label{ggHtbP_self_triangle_diagrams}
\end{center}
\end{figure}
\begin{figure}[]
\begin{center}
\includegraphics[width=0.8\textwidth]{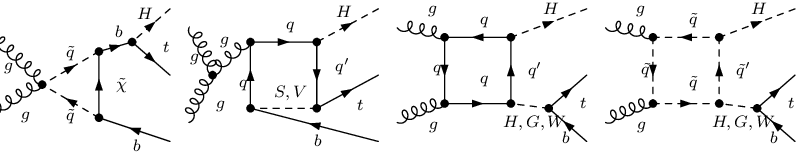}
\includegraphics[width=0.8\textwidth]{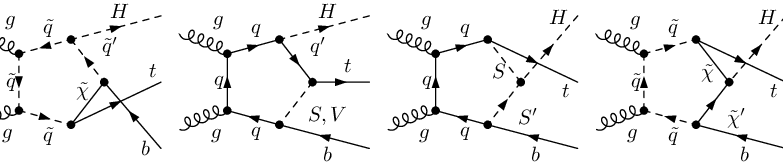}
\caption{Representative diagrams with irreducible four- and five-point vertices. 
$S$ denotes a Higgs or Goldstone boson, $V$  an electroweak gauge boson, $\tilde{q}$ a squark,
and $\tilde{\chi}$ a chargino or neutralino.
}\label{ggHtbP_box_pen_diagrams}
\end{center}
\end{figure}

\begin{figure}[]
\begin{center}
\includegraphics[width=0.6\textwidth]
{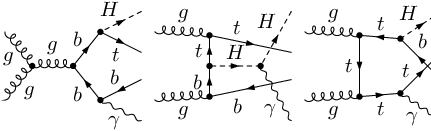}
\caption{Representative diagrams for real photon emission.}
  \label{ggHtb_real_diagrams}
\end{center}
\end{figure}
In this section we discuss the NLO EW contributions to the $\ggHtbm$ subprocess. These corrections are of 
order $\cO(\al_s^2\al^2)$. Other corrections of the same order arising
 from the remaining subprocesses in \eq{eq:qqHtb}, \eq{eq:bbHtb} and \eq{eq:gyHtb} 
are much smaller and will be neglected.  
  
The NLO EW contributions are composed of a virtual part and a real part. 
The virtual part comprises the contributions of bottom-quark and top-quark self-energies,
of triangle, box and pentagon diagrams, and of wave-function
corrections.
For illustration, 
some generic classes of self-energy and vertex diagrams including 
the corresponding counter\-terms are shown in \fig{ggHtbP_self_triangle_diagrams}.
The box and pentagon diagrams are UV finite, a representative sample is depicted  
in~\fig{ggHtbP_box_pen_diagrams}. 

The virtual part contains UV divergences, soft, and collinear singularities.
 The UV divergences are canceled by renormalization, which requires
 the choice of a renormalization scheme.
 We use the same renormalization procedure as the one described in \cite{Dao:2010nu} for 
the process $\bar{b}b \rightarrow W^\mp H^\pm$. 
This is a hybrid of on-shell and $\DRb$ schemes originally defined in \cite{Frank:2006yh}. 
We summarize here the main points and refer to \cite{Dao:2010nu} for more details. 
The calculation is done by using the technique 
of constrained differential renormalization \cite{delAguila:1998nd} which 
is, at one-loop level, equivalent to regularization by dimensional reduction \cite{Siegel:1979wq, Hahn:1998yk}. 
The on-shell scheme is used for the fermion sector, the fine-structure constant,
 and the charged Higgs-boson mass. 
The charged Higgs field and $\tan\beta$ are renormalized in the $\DRb$ scheme. 
Hence, the correct on-shell behavior of the external $H^-$ must be ensured by 
including the finite wave-function renormalization factor \cite{Hollik:2010dh} 
\bea
\sqrt{Z_{H^-H^+}}=1 - \fr{1}{2}\RE\fr{\partial}{\partial p^2}\hat\Sigma_{H^-H^+}(p^2)\big\vert_{p^2=M_{H^\pm}^2},
\eea 
where $\hat\Sigma_{H^-H^+}(p^2)$ is the $H^\pm$ renormalized self-energy, and the mixing of 
$H^-$ with $W$ and charged Goldstone bosons~(see \fig{ggHtbP_self_triangle_diagrams}). 

To make the EW corrections independent of $\ln m_f$ from the light fermions $f\neq t$, 
we use the fine-structure constant at $M_Z$, $\alpha = \alpha(M_Z)$ as an input parameter. 
This means that we have to modify the counterterm according to
\bea
\delta Z_e^{\alpha(M_Z)}&=&\delta Z_e^{\alpha(0)} - \fr{1}{2}\Delta\alpha(M_Z^2),\crn
\Delta\alpha(M_Z^2)&=&\fr{\partial \Sigma_T^{AA}}{\partial k^2}\bigg\vert_{k^2=0}-\fr{\RE\Sigma_T^{AA}(M_Z^2)}{M_Z^2},
\eea
with the photon self-energy from the light fermions only to avoid double counting. 
In the calculation of EW corrections, the couplings in \eq{b_H_couplings_tree} are used. 

Concerning the bottom quark, the 
pole mass enters the kinematical variables of the
matrix element and the phase space, whereas the $b$ Yukawa couplings
are usually improved by using the running $m_b$ (as done \eg in the 
calculation of NLO QCD contributions \cite{Dittmaier:2009np}). 
For NLO EW calculations, however, such a distinction is not
possible since the $b$-quark mass is of EW origin. 
One has to use a common value for the kinematical variables  
and for the Yukawa couplings in order to obtain UV finiteness
because of the interplay between the bottom-mass, 
the bottom-Goldstone and the bottom-Higgs couplings in the 
renormalization of the EW contributions.
The use of different masses would violate
important Ward identities involving $m_b$ 
(see \eg \cite{Baro:2008bg}), 
leading to an incomplete cancellation of UV poles. 
Hence, one can either choose the pole mass or the running mass
in all places. We have decided to take the running mass $m_b = m_b^{\DRb}(\mu_R)$ 
because 
a more accurate treatment of the Yukawa couplings is more significant
than an accurate treatment of the kinematics. 
For infrared-safe observables the kinematical logarithms of $m_b$ cancel. 
For non-infrared-safe observables like in our case 
(see discussion below) some contribution of $\alpha\log(m_b)$ remains. 
Ideally, this would be $\alpha\log(m_b^\text{pole})$. 
The difference $\alpha\log(m_b^{\overline{DR}}/m_b^\text{pole})$ 
is, however, of higher order and numerically very small, and hence
can be neglected in our study. Moreover, if the hadronization of 
the $b$ quark is taken into account, the kinematical $m_b$ dependence 
is expected to be irrelevant and one can regard the kinematical $m_b$ 
as a regulator. 

We classify the virtual part into two gauge-invariant groups. The first group 
consists of one-loop diagrams contributing to the process 
\be
g+g \to H^- + H^{+*} \to H^- + t+\bar{b},\label{eq:ggHH}
\ee
where the virtual $H^{+*}$ can be on-shell, see \fig{ggHtbP_self_triangle_diagrams} and \fig{ggHtbP_box_pen_diagrams} (box 
diagrams). 
The second group is the remainder, which 
is free of resonating propagators. 
The first group is UV and infrared finite since the channel $g+g \to H^- + H^{+*}$ does 
not occur at tree level. Because the intermediate $H^{+*}$ can be on-shell, 
special care has to be taken for the 
numerical integration over the phase space. The resonance propagator
reads
(zero-width approximation)
\bea
\Delta_{H^\pm}=\fr{1}{q^2 - M_{H^\pm}^2 + i\eps}=\text{PV}\left(\fr{1}{q^2 - M_{H^\pm}^2} \right) - i\pi\delta(q^2 - M_{H^\pm}^2),
\eea
where $\text{PV}$ denotes the Cauchy principal value. The principal-value part 
can be calculated by imposing a small cut on $q^2$ around the pole. The contribution from 
the $\delta$ function part is nonvanishing because the imaginary part of 
the on-shell propagator can multiply by the imaginary part of the 
loop integrals, hence the corresponding one-loop 
amplitude can interfere with the tree-level amplitude. We have checked that this contribution 
is indeed nonzero, but small. A naive calculation taking into account only the principal value 
part would lead to an incorrect result. For practical purposes,
a better method is introducing a small width in the resonance propagator,
\bea
\Delta_{H^\pm}&=&\fr{1}{q^2 - M_{H^\pm}^2 + iM_{H^\pm}\Gamma_{H^\pm}}.
\eea
We have checked that the result is practically independent of 
the small values of the width and agrees with the 
sum of the principal value and $\delta$ function contributions. 
We also notice that this method gives smaller integration error. 
As will be shown in the numerical study, the effect of 
the $H^-H^{+*}$ production mechanism is small at the cross section level, but 
is of importance for differential cross sections.

The real EW corrections arise from the photonic bremsstrahlung process, 
\be
g+g \to H^-+ t+\bar{b} + \ga,\label{eq:ggHtb}\ee
with the corresponding Feynman diagrams shown in~\fig{ggHtb_real_diagrams}. 
This contribution is
divergent in the soft limit ($p_{\ga}^0\to 0$) and contains quasi-collinear corrections \cite{Catani:2002hc} 
proportional to $\alpha\log(m_b^2/E_b^2)$, $E_b$ being the $b$-quark energy, 
in the limit $p_bp_\ga\to \cO(m_b^2)$. 
The $b$-quark mass is used for regularization
and to separate the singular terms. 
A fictitious photon mass ($\la_\ga$) is used for regularization of the
soft singularities. 
If we consider the total cross section, 
\ie\ without applying the cuts in~\eq{eq_cuts},
the soft and quasi-collinear singularities cancel completely in the sum of the virtual and the real contributions, 
according to the Kinoshita-Lee-Nauenberg theorem \cite{Kinoshita:1962ur, Lee:1964is}. 
This requires that we have to use $m_b = m_b^{\DRb}(\mu_R)$ as in the virtual amplitudes. 
If the cuts in \eq{eq_cuts} are imposed then the soft singularities still cancel, but the 
quasi-collinear singularities do not, since the cuts requiring bottom-photon separation are not collinear safe. 
In this case, some quasi-collinear singularities remain and are regularized by the bottom mass. 
Those left-over singularities can be separated, as discussed below. 
If a sufficiently collinear $b$-photon system is recombined before applying cuts then 
the quasi-collinear singularities cancel, but the result will depend on the recombination parameter. 
As done in the previous study for the NLO QCD corrections \cite{Dittmaier:2009np}, 
we assume in this paper bottom-photon separation, and hence no photon recombination is applied. 

\begin{figure}[]
 \begin{center}
\includegraphics[width=7cm]{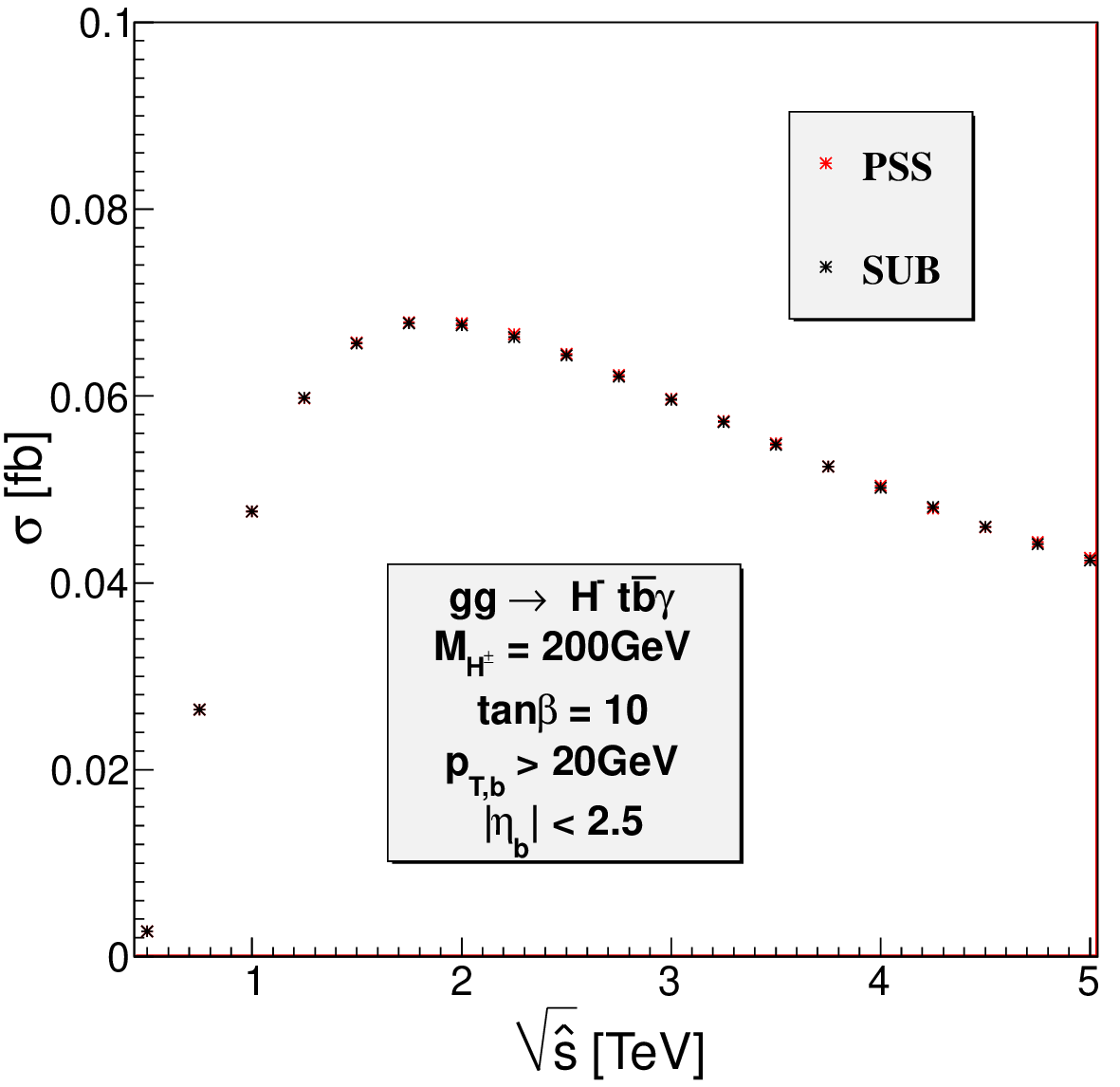}
\includegraphics[width=7cm]{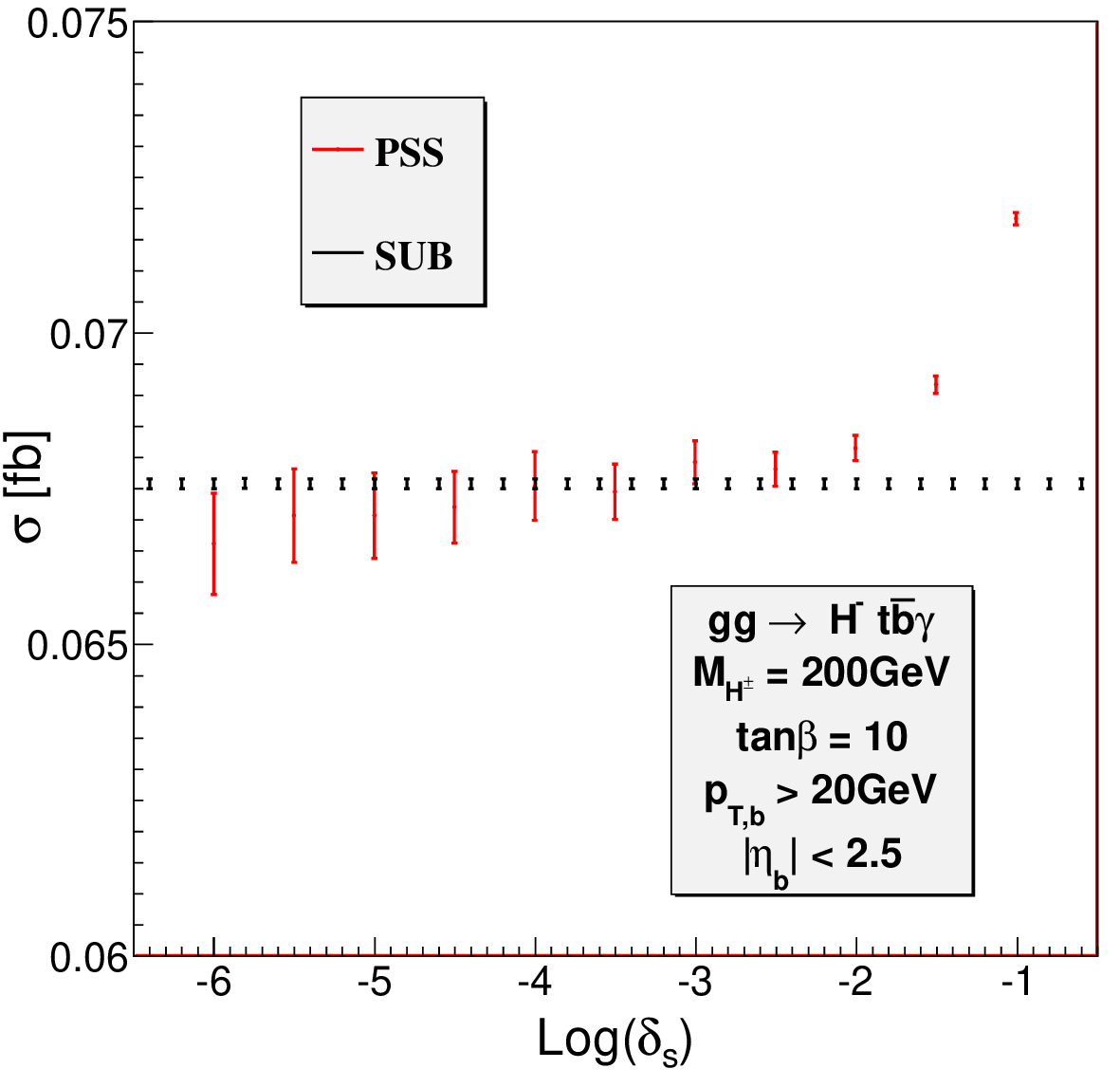}
\caption{Partonic cross sections as functions of the center-of-mass energy $\sqrt{\hat s}$ in the left panel and  of the cutoff parameter $\de_s$ in the right panel are
presented for the $\ggHtby$ process. The results are obtained by using the
phase space slicing method (PSS) and the dipole subtraction method (SUB, straight line) in the CPX scenario. 
The photon mass regulator $\ln \la_\ga$ is set to zero.} \label{fig:PSS_SUB_ggHtby}
\end{center}
\end{figure}

The dipole subtraction method \cite{Catani:1996vz, Dittmaier:1999mb, Catani:2002hc, Dittmaier:2008md} is used to extract 
the singularities from the real corrections and combine them with the virtual contribution. 
The subtraction method for doing the phase-space integration for 
the radiation process \eq{eq:ggHtb} 
arranges the integral in the following way,
\begin{equation}
  \label{eq:dipmaster}
  \sigma_{\text{real}} = \int_4 \left[ \,
    d\sigma_\text{real}\;\theta(p_b) -  d\sigma_\text{sub}\; \theta(\tilde{p}_b)\, \right]
  + \int_4 d\sigma_\text{sub} \; \theta(\tilde{p}_b).
\end{equation}
The subscript $4$ refers to the 4-body final state including the radiated
photon, $\theta$ is a function to impose the kinematical cuts defined in~\eq{eq_cuts}, 
$\tilde{p}_b$ is a function 
of $p_i$ with $i=$ $H^{-}$, $t$, $b$, $\gamma$, with the definition given 
in~\cite{Dittmaier:1999mb, Catani:2002hc, Dittmaier:2008md}. 
The subtraction function $d\sigma_\text{sub}\theta(\tilde{p}_b)$ has to be 
chosen such that the first integral is finite and the second one can be partially 
analytically integrated over the singular variables. The function $d\sigma_\text{sub}$ 
has the same singular structure as $d\sigma_\text{real}$ pointwise in the phase space.  
There are two ways to deal with the cut function.\\ 
\noindent i) We require that (the pseudorapidity cut is neglected to simplify the discussion)
\bea
\theta(\tilde{p}_b) \to \theta(p_b) \quad \text{or}\quad \tilde{p}_b \to p_b
\label{eq_cut_limit}
\eea
in the singular limits (the soft limit is 
trivially satisfied), which implies that $\theta(\tilde{p}_b)$ is not collinear safe, 
so that the first integral is soft and (quasi-)collinear finite.  
All soft and quasi-collinear singularities are contained in the second integral. 
All soft and some quasi-collinear singularities are canceled in the sum with the virtual contribution. 
The leftover quasi-collinear singularities, regularized by $m_b$, 
can be factorized and separated. A detailed procedure including the 
definition of $\tilde{p}_b$ is described in \cite{Dittmaier:2008md}. 
A consequence of the condition~(\ref{eq_cut_limit}) is that, in the calculation of the first integral, 
we can set $m_b = 0$ in the kinematics (but not in the Yukawa couplings).\\
\noindent ii) We require that the cut function $\theta(\tilde{p}_b)$ 
is infrared safe as in \cite{Dittmaier:1999mb, Catani:2002hc} so that the sum of the second integral 
and the virtual contribution is independent of soft and quasi-collinear singularities. Specifically, it means 
that the condition~(\ref{eq_cut_limit}) is satisfied for the soft limit but not for the collinear limit. 
The first integral, therefore, contains the leftover quasi-collinear singularities. 
Since the result is finite one can do it numerically. In this approach, one has to 
keep $m_b$ everywhere. 
We have implemented 
both approaches and found good agreement for the cross section and the distributions. 
Moreover, the result of the dipole subtraction method is compared with the one of the phase-space slicing method, as illustrated 
in \fig{fig:PSS_SUB_ggHtby}. In the numerical analysis, we will present the
results of the dipole subtraction method because the integration errors are smaller.

The above treatment of the kinematical cuts in the dipole subtraction
method is also applied for the bottom-quark histograms displayed in
Section~\ref{sec:distributions}.

Finally, the hadronic cross section at NLO is written in the following
way as the sum of the improved Born approximation and the 
genuine loop and radiation terms for the $gg$ subprocess,
\bea
\sigma^{pp}_{\text{NLO}}&=&\sigma^{pp}_{\text{IBA}} +
\Delta^{pp/gg}_{\text{EW}} \, ,
\eea
 with
\bea
\Delta^{pp/gg}_{\text{EW}}  &=& 
\int \dd x_1\dd x_2 \; F_g^{p}(x_1, \mu_F)F_g^{p}(x_2, \mu_F) \;
\Delta^{gg}_{\text{EW}}(\alpha_s^2\alpha^2,\mu_R) \, .
\eea
Thereby, $\De_{\text{EW}}^{gg}$ is the sum of  
the virtual and the real corrections at the partonic level,
as discussed above. 
The IBA part of the cross section results from the various sources at
the partonic level, 
\bea
\sigma^{pp}_{\text{IBA}}  &=& \sigma^{pp/gg}_{\text{IBA}} 
+ \sigma^{pp/q\bar{q}}_{\text{IBA}} 
+ \sigma^{pp/b\bar{b}}_{\text{IBA}} 
+ \sigma^{pp/g\gamma}_{\text{IBA}} \, ,
\eea
as discussed in Section~\ref{sec:LO_ggHtb}.

\section{Numerical studies}
\label{sect-results}

\subsection{Input parameters}
\label{sect-input}
We use the same set of input parameters as in \cite{Dao:2010nu} for the sake of comparison. For the SM sector: 
\begin{equation}
\begin{aligned}
\alpha_s(M_Z) &= 0.1197, \hs &\alpha(M_Z)&=1/128.926, \\
M_{W} &= 80.398\gev, \hs& M_Z&= 91.1876\gev, \\
m_t & =173.1\gev, \hs &\mbmb& = 4.2\gev.
\end{aligned}
\end{equation} 
We take here $\alpha_s = \alpha_s^{\MSb}(\mu_R)$ at three-loop order \cite{Amsler:2008zzb}.
 $\mbmb$ is the QCD-$\MSb$ $b$-quark mass, while the top-quark mass is understood 
as the pole mass. 
The CKM matrix is approximated to be unity.

For the soft SUSY-breaking parameters, the adapted CP-violating benchmark scenario (CPX)~\cite{Williams:2007dc,Carena:2000ks} is used,
\bea\beal 
|\mu| &= 2\tev, |M_2|=200\gev,\, |M_3| = 1\tev,\, |A_t|=|A_b|=|A_\tau|=900\gev,\\
M_{\tilde Q}&=M_{\tilde D}=M_{\tilde U}=M_{\tilde L}=M_{\tilde
  E}=M_{\text{SUSY}}=500\gev, \, |M_1|= 5/3\tan^2\theta_W |M_2|.
\eeal\eea
We set $A_f=0$ for $f=e,\mu,u,d,c,s$ since the Yukawa couplings of the first two fermion generations proportional to the small fermion masses are neglected in our calculations. 
With the convention that $M_2$ is real, 
the complex phases of the trilinear couplings $A_t$, $A_b$, $A_\tau$ and the gaugino-mass parameters $M_1$ and $M_3$ are chosen as default according to
\bea
\phi_{t}=\phi_{b}=\phi_{\tau}=\phi_{1}=\phi_{3}=\fr{\pi}{2},
\eea
unless specified otherwise. 
The phase of $\mu$ is chosen to be zero. This is consistent with the experimental data  of
the electric dipole moment and the explanation of the muon anomalous magnetic moment discrepancy 
between the present data and the standard model prediction (see \eg\cite{Stockinger:2006zn}). 
We will study the dependence of our results on $\tan\beta$, $M_{H^\pm}$ and $\phi_t$ in the numerical analysis. 

The scale of $\al_s$ in the SUSY-QCD resummation of the effective bottom-Higgs couplings in 
Eq.~(9) of \cite{Dao:2010nu} is set to be $Q=(m_{\tilde{b}_1} + m_{\tilde{b}_2} + m_{\tilde{g}})/3$. 
This choice is justified by the two-loop results for the $\Delta_b$ corrections \cite{Noth:2008tw, Noth:2010jy, Mihaila:2010mp}. 
If not otherwise
specified, we set the renormalization scale equal to the factorization scale, $\mu_R=\mu_F$, in all numerical
results. Our default choice for the factorization scale is $\mu_{F0} = (m_t + M_{H^\pm})/3$, which 
is justified by the NLO QCD results \cite{Dittmaier:2009np}. 
We use the MRST2004qed code to calculate the PDFs.

Our study is done for the LHC at $7\tev$, $8\tev$ and $14\tev$ center-of-mass energy ($\sqrt{s}$). 
In the following we show the dependence of the cross section on $\tan\beta$, $M_{H^\pm}$ and $\phi_t$, and 
various differential distributions for the default parameter point. 
Since the results of the different center-of-mass energies look quite similar in shape
and differ mainly by the magnitude of the cross section, our discussion essentially applies
to all displayed cases of the total energy.

\subsection{Calculations and checks}
\label{sect-ggWH-qcd-gauge}
The results in this paper have been obtained by two independent calculations. 
We have produced, with the help of \fav \cite{Hahn:2000kx, Hahn:2001rv} and \fcv \cite{Hahn:1998yk}, two different Fortran~77 codes. 
The loop integrals contain five-point tensor integrals up to rank
three, and four-point tensor integrals up to rank three. 
The pentagon integrals are reduced to the box 
integrals by using the reduction methods in \cite{Denner:2002ii, Binoth:2005ff, Denner:2005nn}. 
The two-, three- and four-point tensor integrals are further reduced
to the scalar integrals by using the 
Passarino-Veltman reduction method~\cite{Passarino:1978jh}. 
The loop integrals are evaluated with two independent libraries, 
\ltff \cite{looptools,vanOldenborgh:1989wn,looptools_5p} using the 
five-point reduction method of \cite{Denner:2002ii} and our in-house 
library \lis\ using the 
five-point reduction method of \cite{Denner:2005nn}. 
The latter uses the method of \cite{hooft_velt, Nhung:2009pm, Denner:2010tr}, treats all the internal 
masses as complex parameters and has an option to use quadruple precision, on the 
fly, when numerical instabilities are detected. 
The phase-space integration is done by using the Monte Carlo
integrators \bases~\cite{bases} and \vegas~\cite{Lepage:1977sw}. 
The results of the two codes are in good agreement. 
On top, we have also performed a number of other checks: 

For the process $\ggHtbm$, we have verified that the results are QCD gauge invariant at LO, 
IBA and NLO. This nontrivial check, which can detect a bug in the Feynman rules and in 
the tensor reduction procedure, can be easily done in practice by changing the numerical value of 
the gluon polarization vector $\eps_{\mu}(p,q)$, where $p$ is the gluon momentum and $q$ is 
an arbitrary reference vector. QCD gauge invariance means that the squared amplitudes are independent 
of $q$. More details can be found in \cite{Boudjema:2007uh}. 
The common checks of UV and infrared finiteness are done for the NLO calculations. 

\subsection{LO, IBA and NLO cross sections}
\begin{figure}[t]
 \begin{center}
\includegraphics[width=7cm]{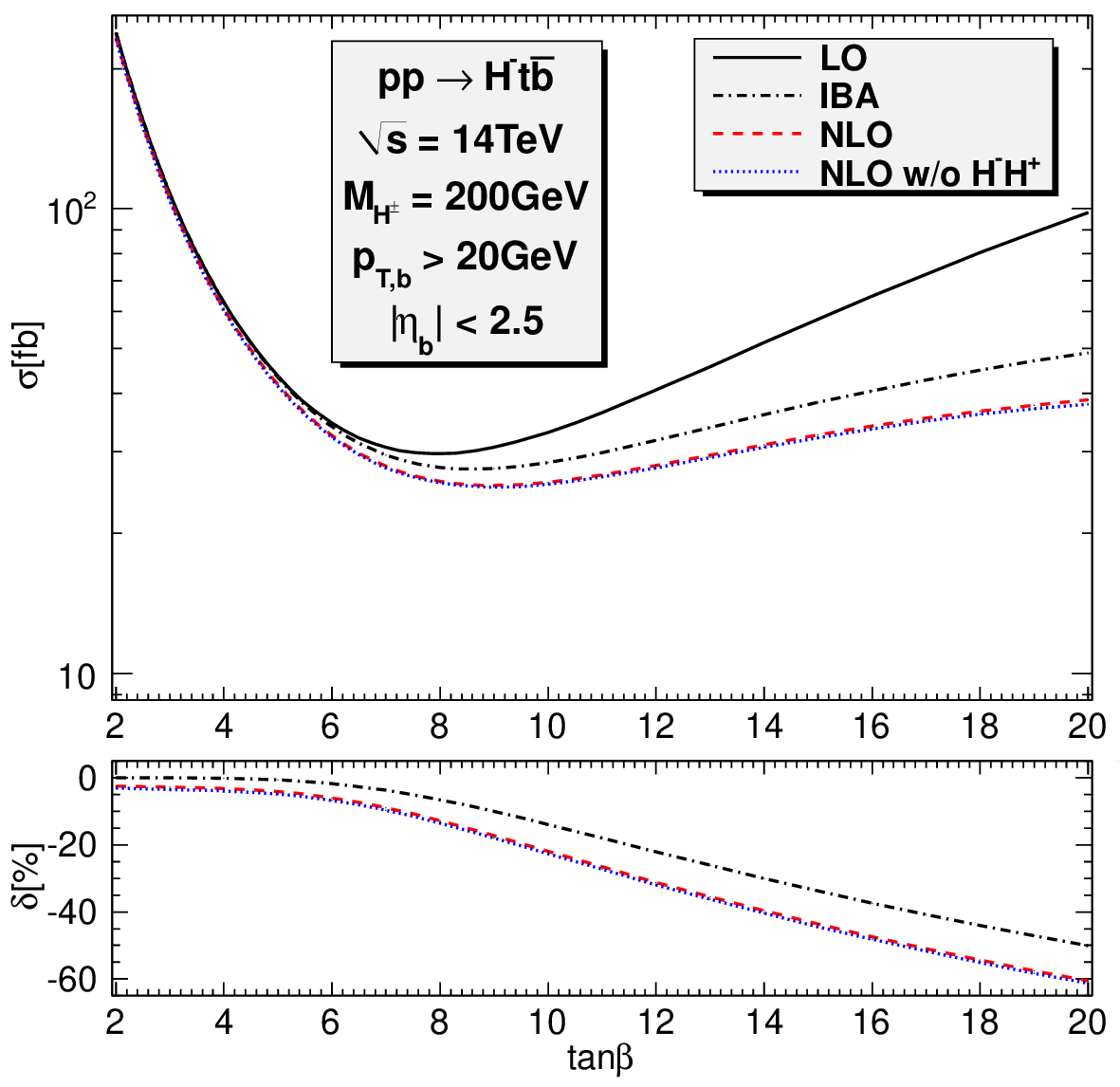}
\includegraphics[width=7cm]{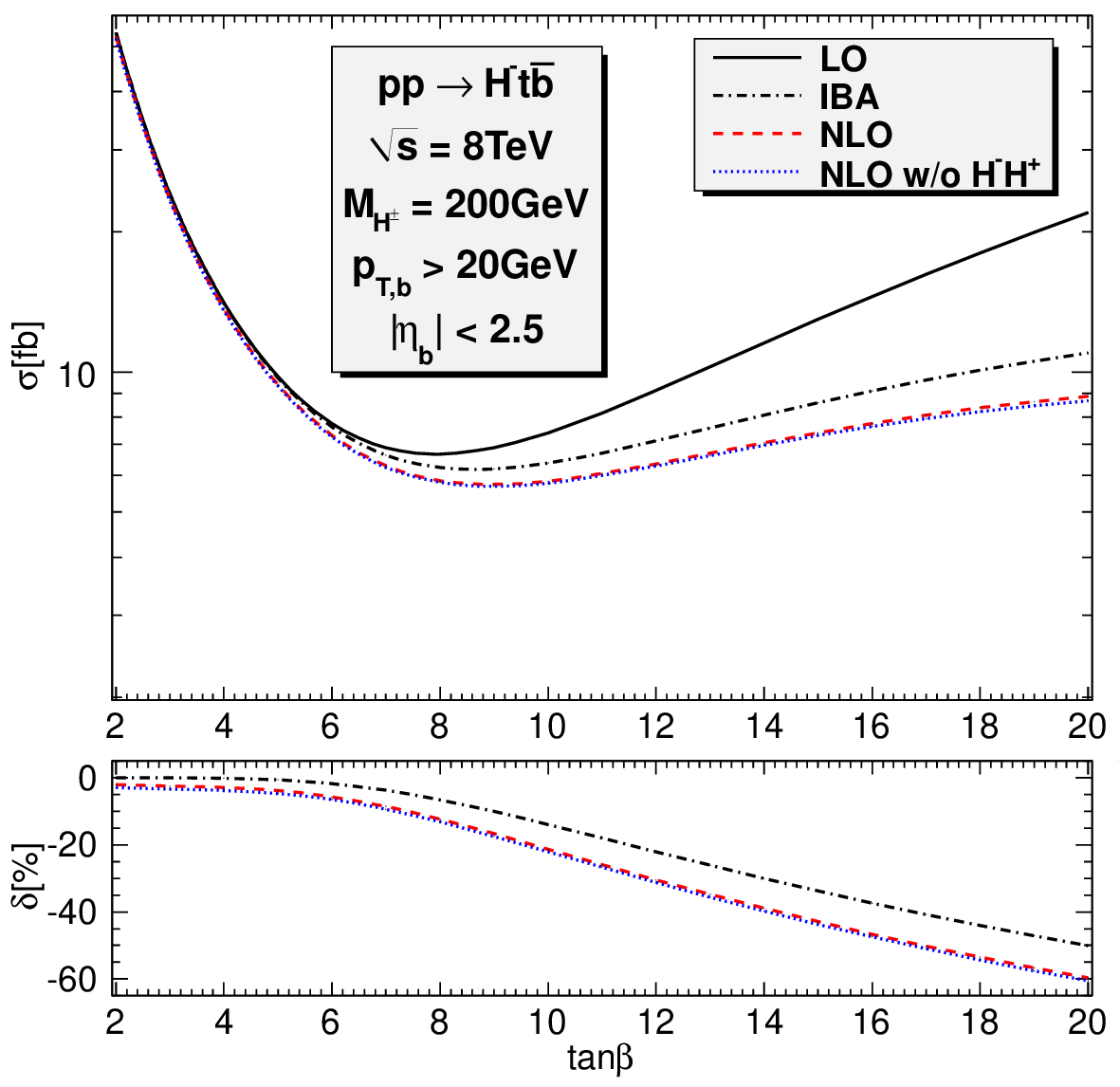}
\includegraphics[width=7cm]{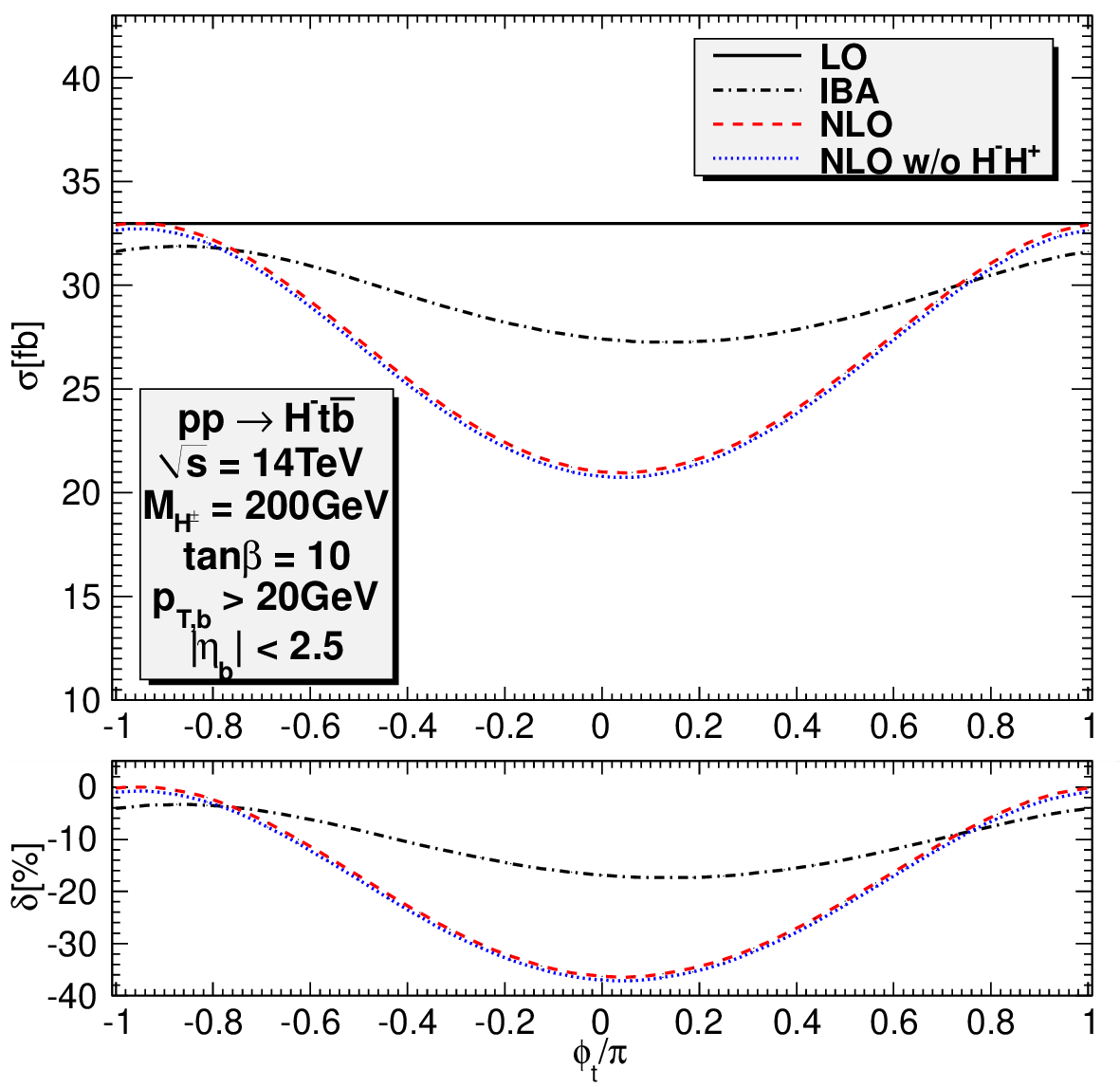}
\includegraphics[width=7cm]{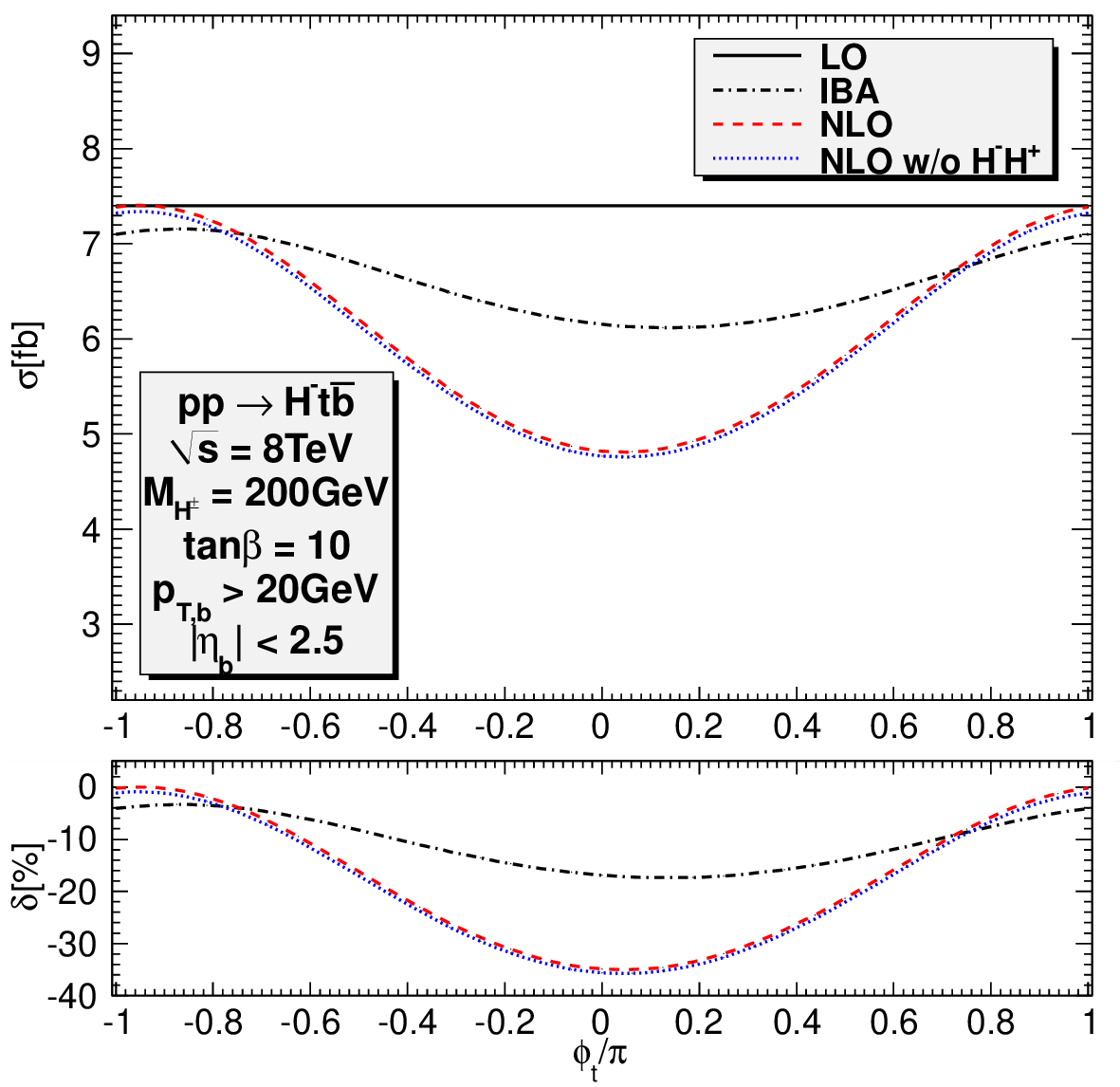}
\includegraphics[width=7cm]{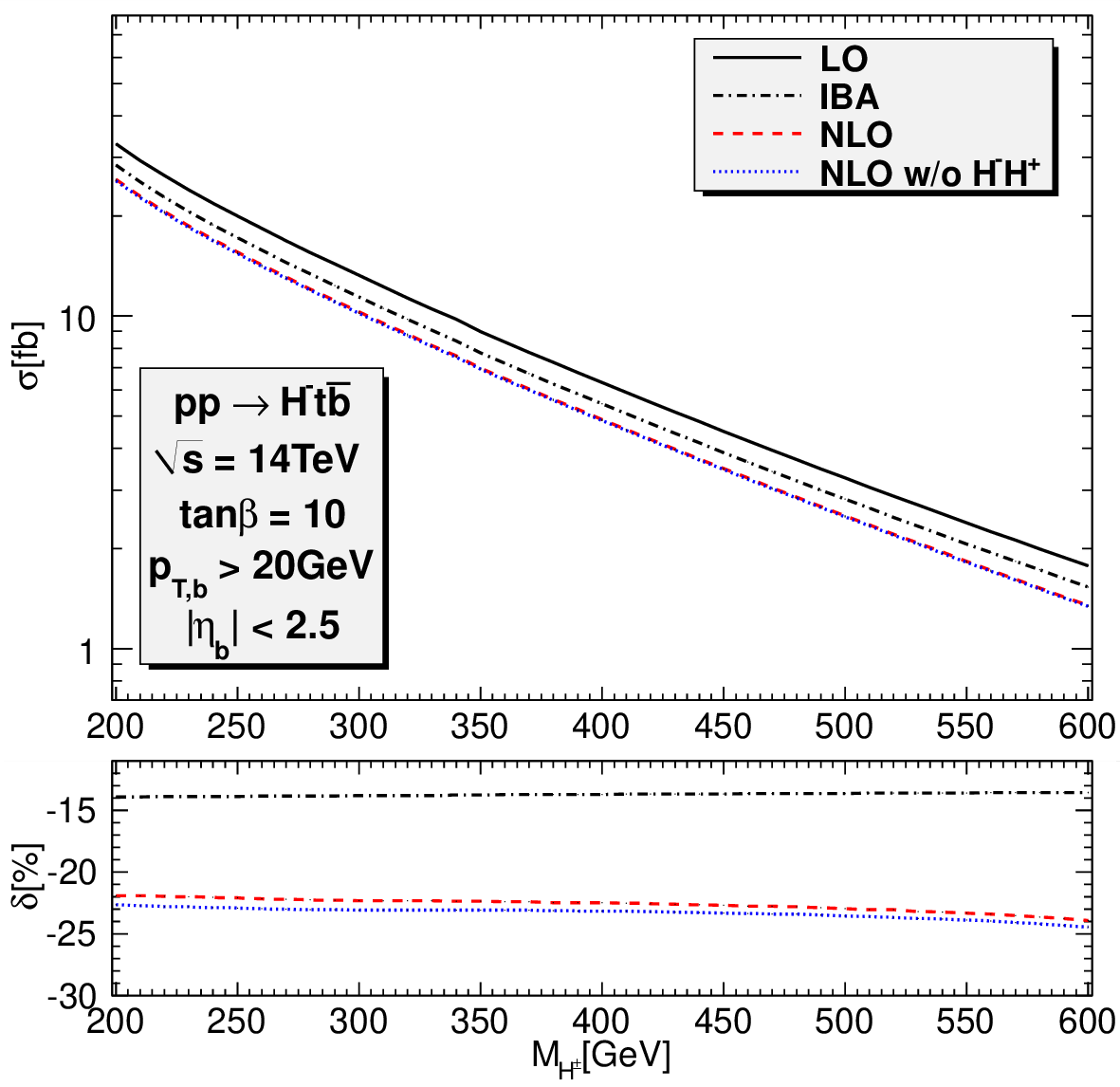}
\includegraphics[width=7cm]{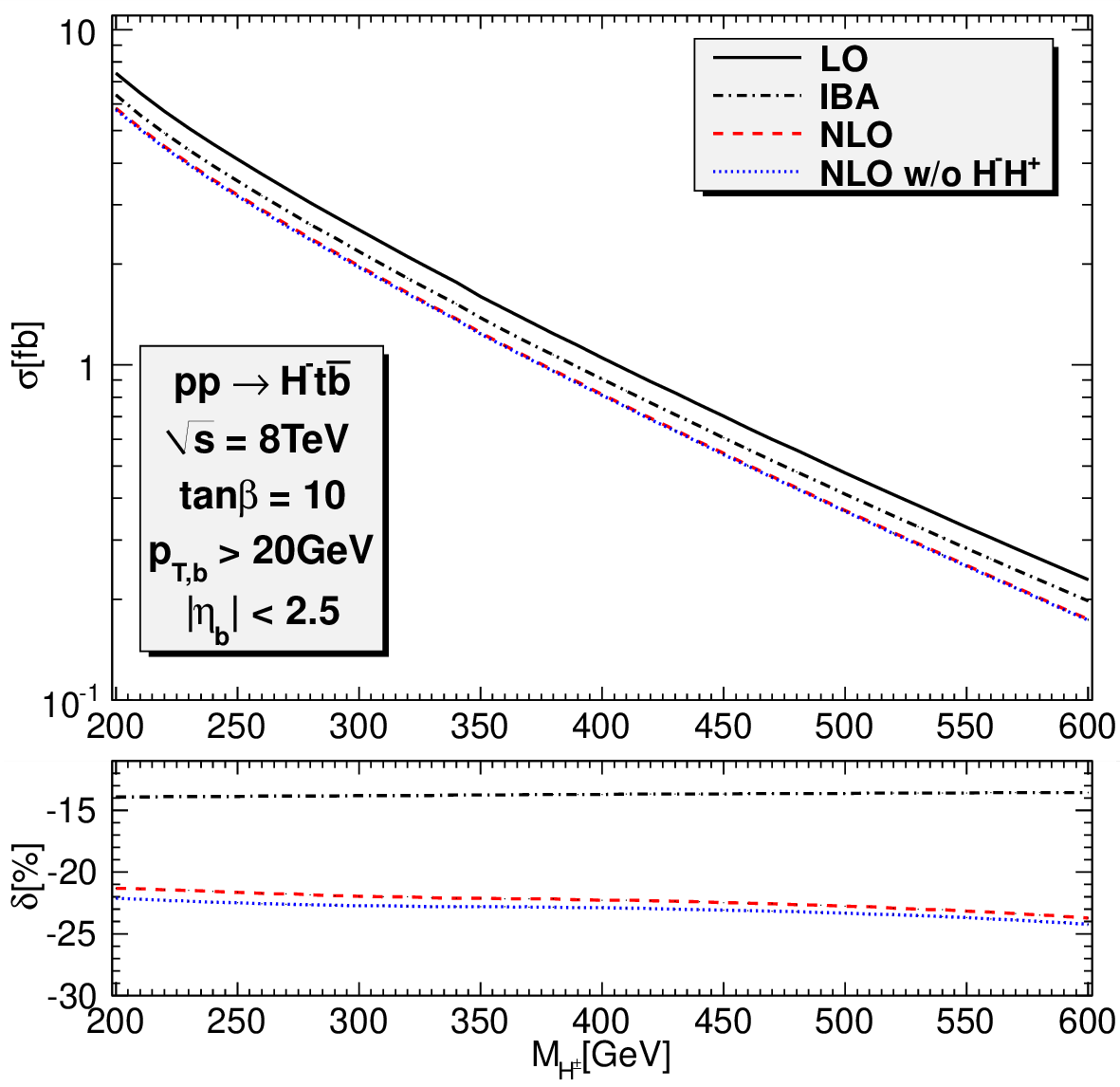}
\caption{
The cross section for $pp\to  H^-t\bar b$ as function of
$\tan\beta$, $\phi_t$, $\MHpm$ 
in various steps of approximation LO, IBA, and NLO, for  
$\sqrt{s}=14\tev$ (left) and 
$\sqrt{s}=8\tev$ (right). 
The lower part of each panel shows the relative corrections to the LO results. 
Also shown is the NLO result without the $H^-H^{+*}$ production mechanism.}
\label{fig:Xsection_ppHtb}
\end{center}
\end{figure}

\begin{figure}[t]
 \begin{center}
\includegraphics[width=7cm]{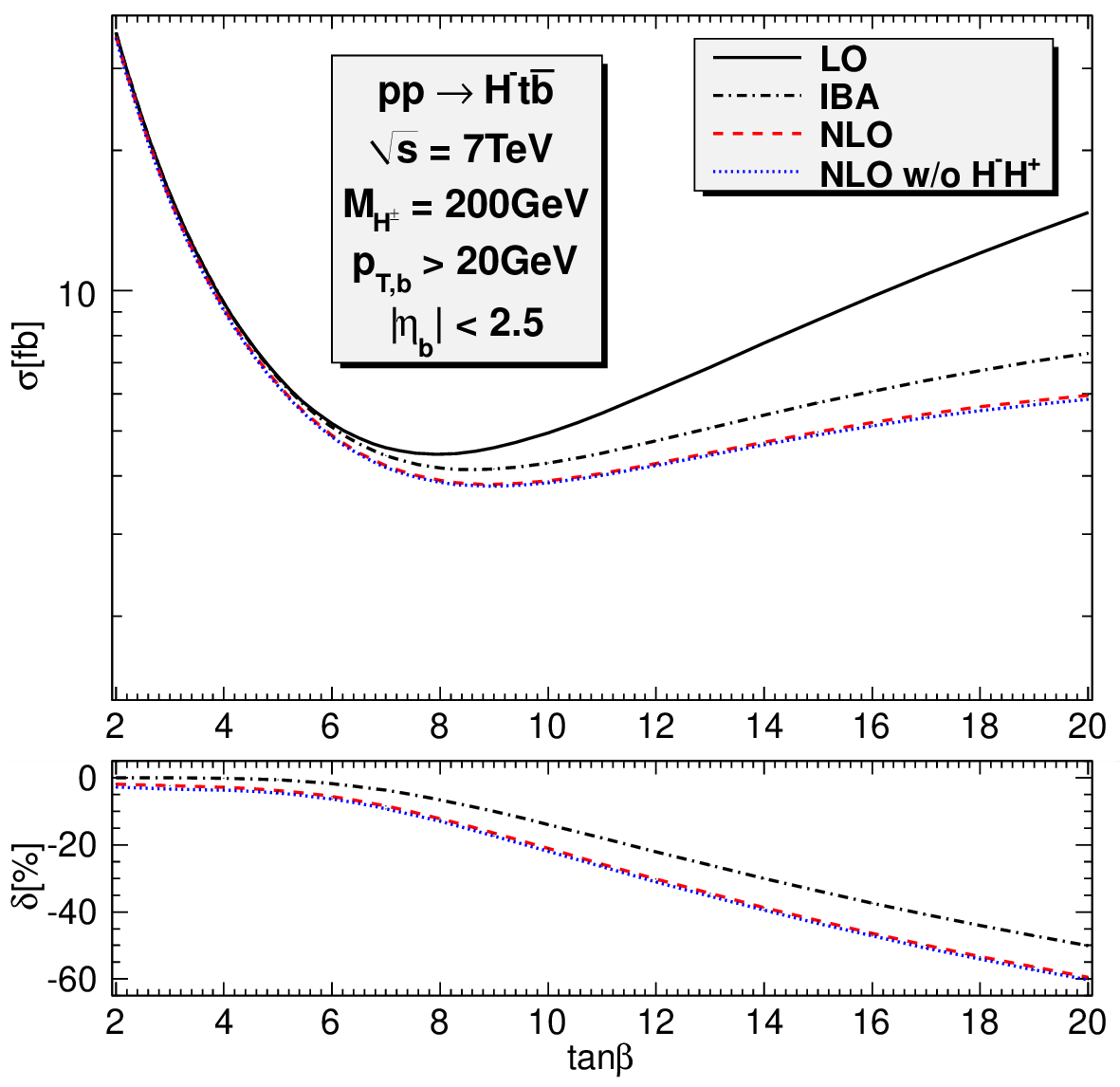}
\includegraphics[width=7cm]{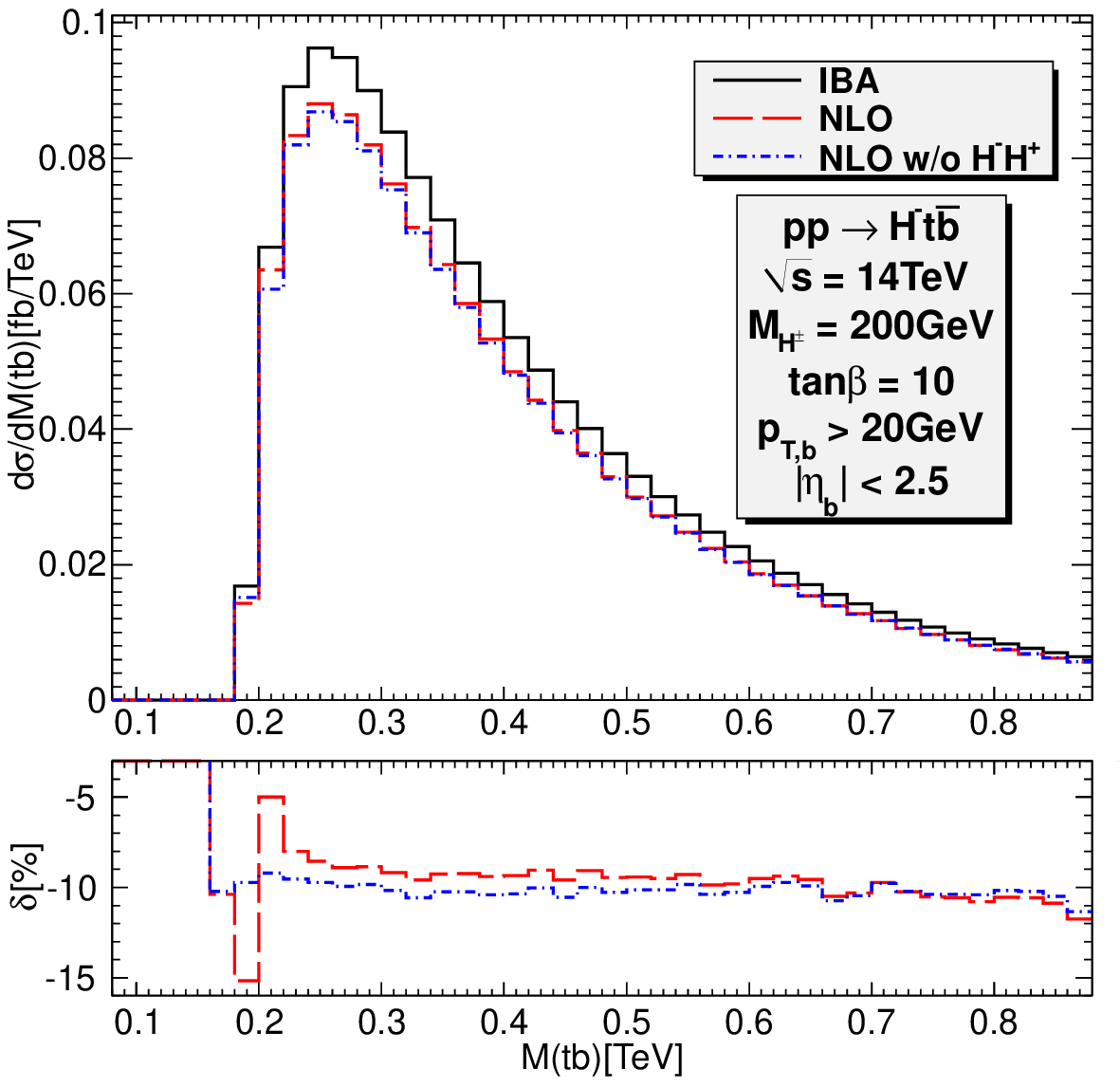}
\includegraphics[width=7cm]{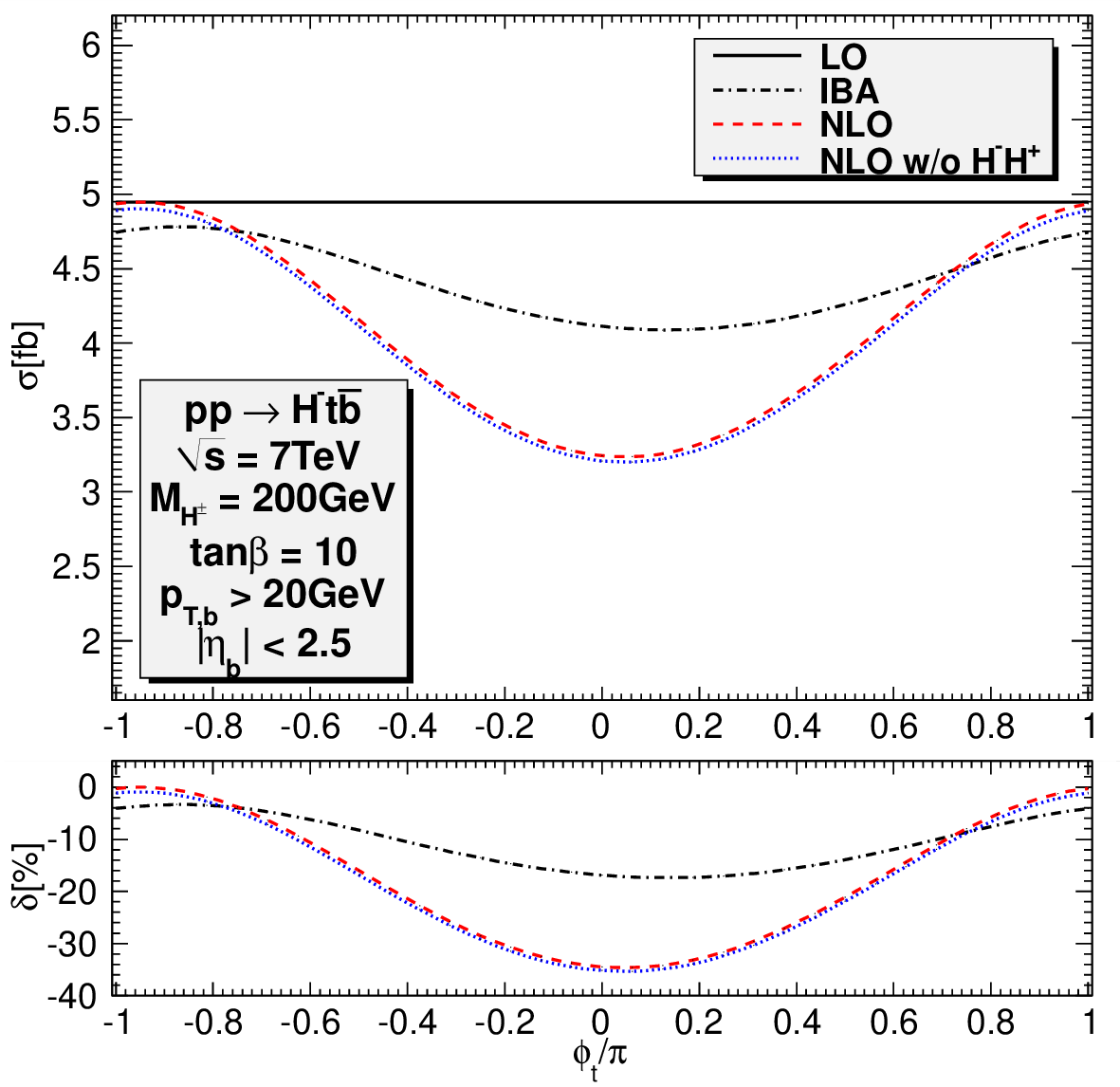}
\includegraphics[width=7cm]{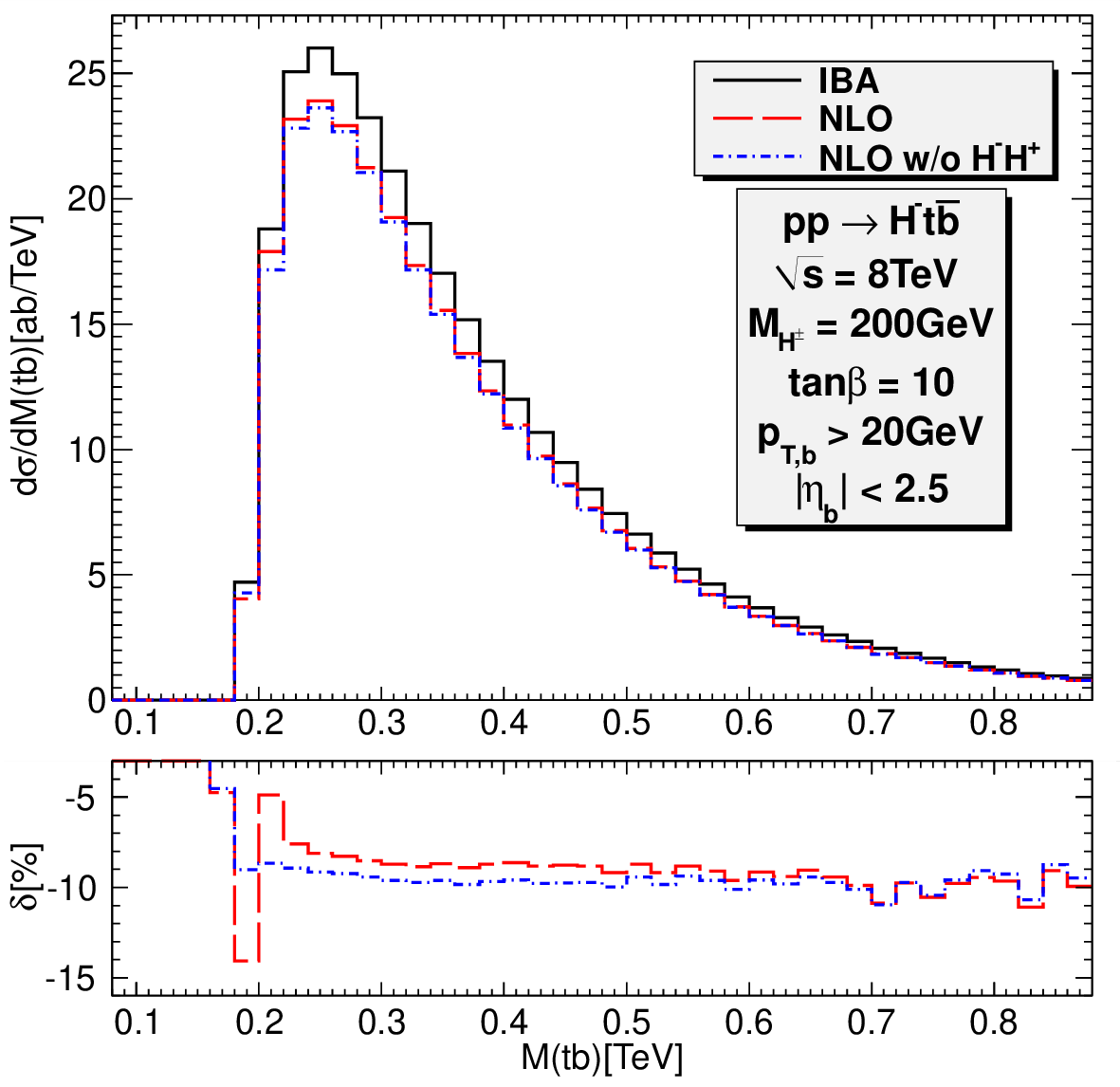}
\includegraphics[width=7cm]{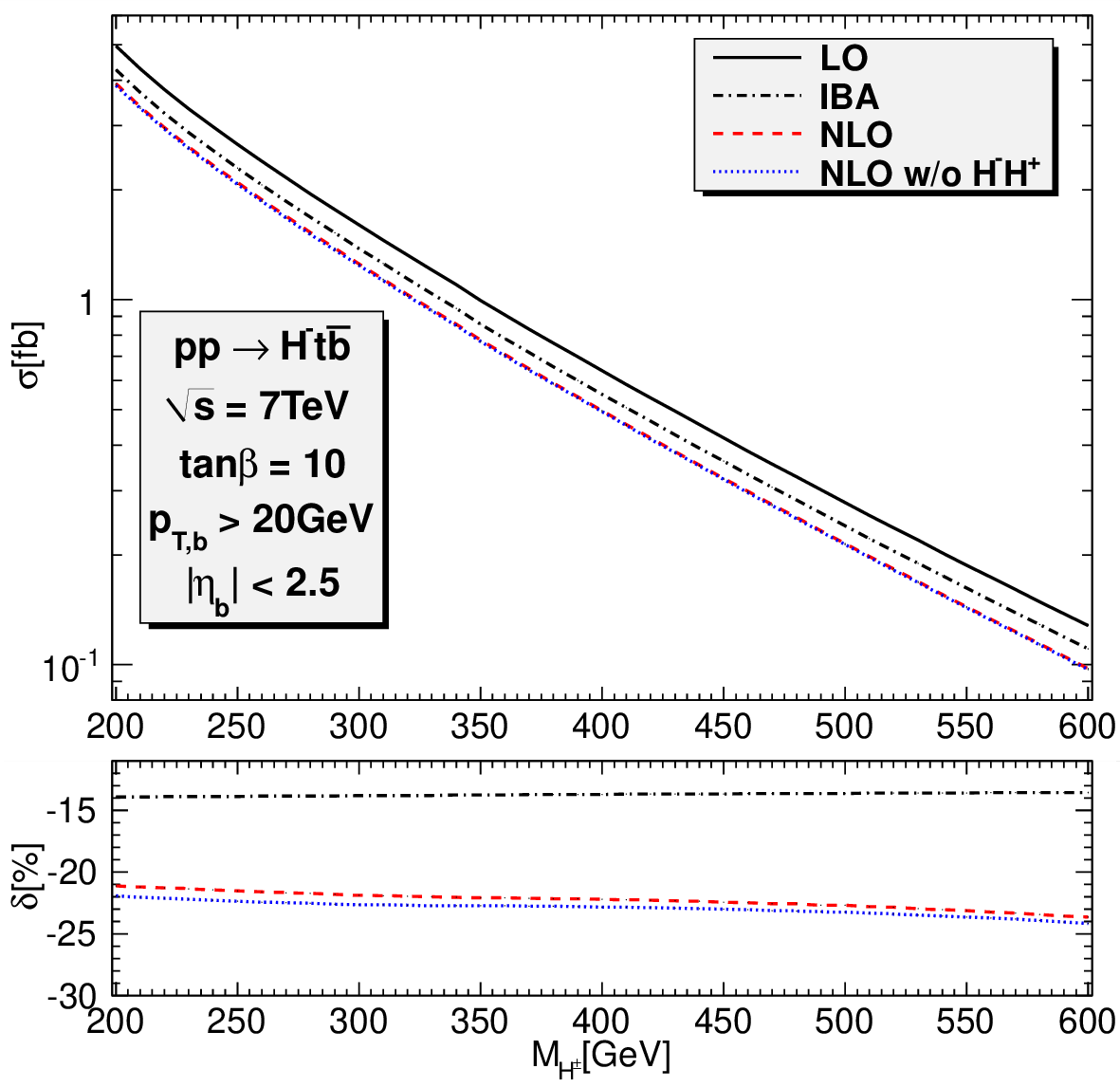}
\includegraphics[width=7cm]{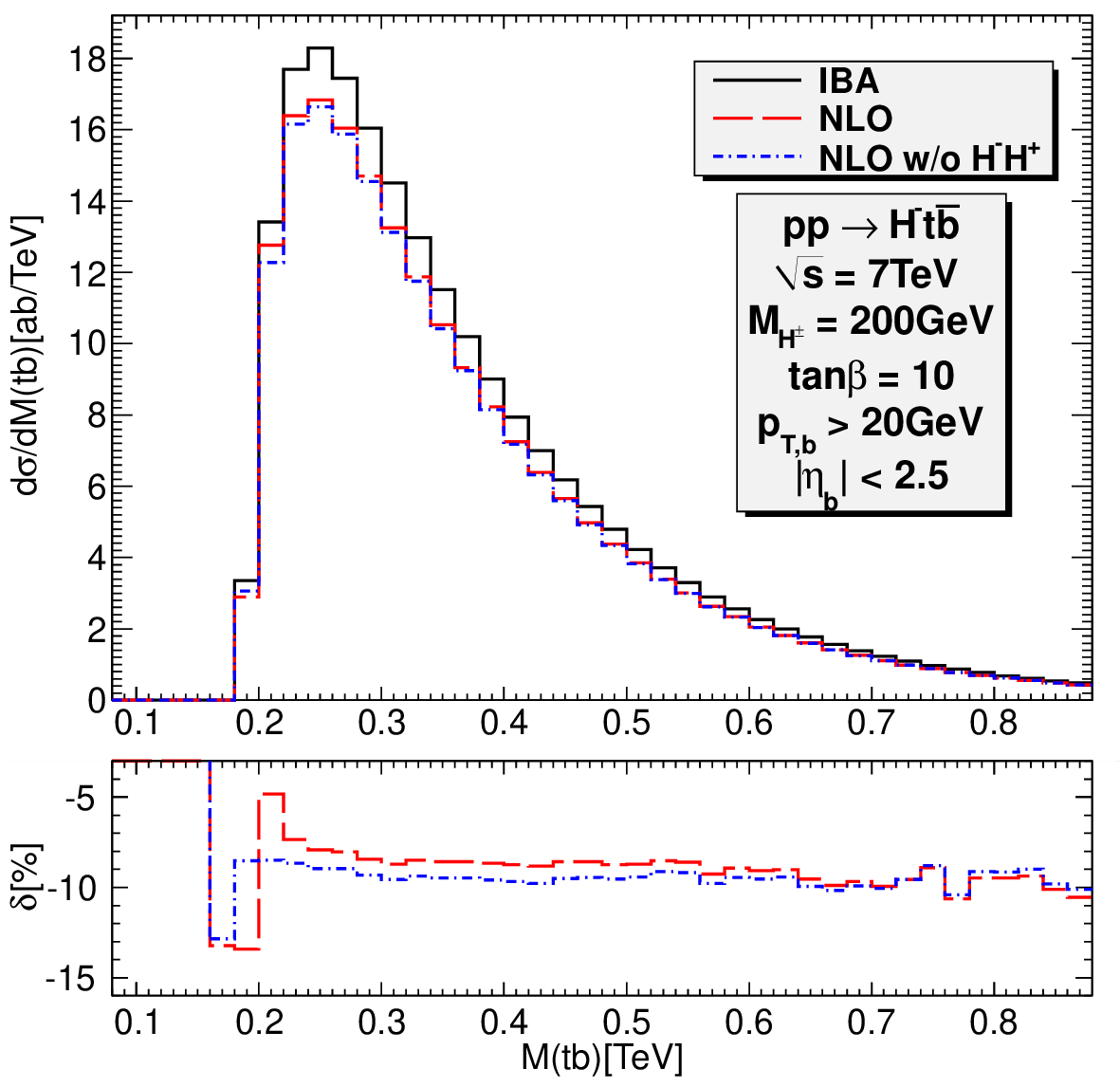}
\caption{Left column: the LO, IBA and NLO cross section for $pp\to  H^-t\bar b$ as function of
$\tan\beta$, $\phi_t$, and $\MHpm$, for $\sqrt{s}=7\tev$. 
Right column: 
the IBA and NLO EW invariant mass distributions of the $t\bar{b}$ system
for $\ppHtbm$ at $14$, $8$ and $7\tev$. 
The lower panels show the relative corrections. 
Also shown is the NLO result without the $H^-H^{+*}$ production
mechanism.}
\label{fig:dist_m45}
\end{center}
\end{figure}

The LO, IBA and NLO cross sections as functions of $\tan\beta$, the phase $\phi_t$, and $M_{H^\pm}$ 
are shown at $14\tev$ and $8\tev$ on the left and right columns of \fig{fig:Xsection_ppHtb}, respectively, and 
at $7\tev$ on the left column of \fig{fig:dist_m45}. 
The relative corrections $\de$, with respect to the LO cross section,
are defined as $\de = (\si_{\text{IBA/NLO}} -\si_{\text{LO}})/\si_{\text{LO}}$. 

We first study  the effects of $\Delta_b$ resummation in the effective bottom-Higgs couplings. 
For small values of $\tan\beta$ the left-chirality  contribution proportional to $m_t/\tan\beta$ is dominant while 
the right-chirality contribution proportional to $m_b\tan\beta$ dominates at large $\tan\beta$. The cross section 
has a minimum around $\tan\beta = 8$. 
The effect of $\De_b$ resummation is best understood by comparing the phase-dependence plot and 
the others. The important point is that $\De_b$ is a complex number and only its real part can interfere with the LO amplitude. 
Thus, the $\De_b$ effect is minimum at $\phi_{t}=\pm \pi/2$ where the dominant contributions are purely imaginary and is largest at $\phi_{t}=0,\pm \pi$. 
The phase-dependence plot shows that 
the $\De m_b^{\text{SEW}}$ effect can be more than $10\%$. 
From the $\tan\beta$ dependence plots where $\De_b$ is mostly imaginary we see the effect of 
order $\cO(\De_b^2)$, which is about $-15\%$ at $\tan\beta=10$. 

We now turn to the NLO cross sections, which include 
the complete EW corrections to 
the process $\ggHtbm$. 
\fig{fig:Xsection_ppHtb} also contains the effect of the $H^- H^{+*}$ resonance mechanism, which  
is almost invisible at the cross section level. 
The NLO cross section depends strongly on $\phi_t$. 
The IBA results are always closer to the NLO values rather than to the LO ones. 
In particular, for the phase dependence, the IBA shows the qualitative features
of the NLO prediction while the LO cross section is a constant. 
After subtracting 
the $\De m_b^{\text{SEW}}$ corrections, the remaining NLO EW contributions are still sizeable.
The relative correction increases with $\tan\beta$ and $\MHpm$ for the default
 value $\phi_t=\pi/2$;
for fixed default values of $\tan\beta$ and $\MHpm$, it 
is maximal (about $40\%$) at $\phi_t=0$. 
As an aside, we remark that
the IBA and NLO EW effects for 
the process $\ggHtbm$ are similar to the ones found in the $\bbWHp$ study~\cite{Dao:2010nu}. 
The hierarchy of the LO, IBA and NLO contributions is also the same as the one 
found in \cite{Kniehl:2010ea}.
 
\begin{table}[h]
 \begin{footnotesize}
 \bc 
 \caption{\label{table_ppHtb1}{\footnotesize{ 
 The total cross section in fb for $pp \to H^-t\bar b$ including the IBA of the four 
subprocesses
and EW NLO corrections to  $\ggHtbm$  at $\sqrt{s}= 14\tev$.
The charged Higgs-boson masses are given in GeV. The numbers in
brackets show the integration uncertainty in the last digit 
when they are significant.}}}
\vspace*{0.5cm}
\begin{tabular}{l c r@{.}l r@{.}l r@{.}l r@{.}l r@{.}l r@{.}l r@{.}l}
 \hline
$\tan\beta$
& $M_{H^\pm}$ 
&\multicolumn{2}{c}{ $\si_{\text{IBA}}^{pp/gg}$}
&\multicolumn{2}{c}{ $\si_{\text{IBA}}^{pp/q\bar q}$}
&\multicolumn{2}{c}{$\si_{\text{IBA}}^{pp/b\bar b}$}
&\multicolumn{2}{c}{$\si_{\text{IBA}}^{pp/g \ga}$}
&\multicolumn{2}{c}{$\De_{\text{EW}}^{pp/gg}$}
&\multicolumn{2}{c}{all}\\
\hline \hline
5  & 200 & 38&833(7) & 3&581  &  0&319 & 0&559   & -1&522(5) & 41&770(8)  \\ 
10 & 200 & 25&447(5) & 2&372  &  0&210 & 0&367   & -2&642(4) & 25&754(6) \\
20 & 200 & 43&992(8) & 3&973  &  0&357 & 0&630   & -10&24(1) & 38&71(1)\\
10 & 300 & 10&740(2) & 0&457  &  0&075 & 0&139   & -1&126(2) & 10&285(3)\\
10 & 400 &  5&207(1) & 0&143  &  0&031 & 0&064   & -0&556(1) & 4&889(2) \\
10 & 600 &  1&4829(3)& 0&0244 &  0&0069& 0&0183  & -0&1842(3)& 1&3483(5) \\ 
\hline 
\end{tabular}\ec
 \end{footnotesize}
\end{table} 
\begin{table}[h]
 \begin{footnotesize}
 \bc 
 \caption{\label{table_ppHtb2}{\footnotesize{
 Similar to \tab{table_ppHtb1} but for $\sqrt{s}= 8\tev$.} }}
\vspace*{0.5cm}
\begin{tabular}{l c r@{.}l r@{.}l r@{.}l r@{.}l r@{.}l r@{.}l r@{.}l}
 \hline
$\tan\beta$
& $M_{H^\pm}$ 
&\multicolumn{2}{c}{ $\si_{\text{IBA}}^{pp/gg}$}
&\multicolumn{2}{c}{ $\si_{\text{IBA}}^{pp/q\bar q}$}
&\multicolumn{2}{c}{$\si_{\text{IBA}}^{pp/b\bar b}$}
&\multicolumn{2}{c}{$\si_{\text{IBA}}^{pp/g\ga}$}
&\multicolumn{2}{c}{$\De_{\text{EW}}^{pp/gg}$}
&\multicolumn{2}{c}{all}\\
\hline \hline
5  & 200 & 8&197(2)   & 1&314   &  0&051   & 0&151  & -0&315(1)  & 9&399(2)   \\ 
10  & 200 & 5&369(1)  & 0&871   &  0&034   & 0&099  & -0&548(2)  & 5&826(2)   \\ 
20  & 200 & 9&295(2)  & 1&456   &  0&058   & 0&171  & -2&115(7)  & 8&864(8)   \\ 
10  & 300 & 1&9970(6) & 0&1377  &  0&0101  & 0&0332 & -0&2056(8) & 1&9724(10) \\ 
10  & 400 & 0&8535(2) & 0&0361  &  0&0035  & 0&0137 & -0&0900(3) & 0&8169(4)  \\ 
10  & 600 & 0&18947(5)& 0&00444 &  0&00056 & 0&00315& -0&02328(8)& 0&17435(10) \\ 
\hline 
\end{tabular}\ec
 \end{footnotesize}
\end{table} 

\begin{table}[h]
 \begin{footnotesize}
 \bc 
 \caption{\label{table_ppHtb3}{\footnotesize{
 Similar to \tab{table_ppHtb1} but for $\sqrt{s}= 7\tev$.} }}
\vspace*{0.5cm}
\begin{tabular}{l c r@{.}l r@{.}l r@{.}l r@{.}l r@{.}l r@{.}l r@{.}l}
 \hline
$\tan\beta$
& $M_{H^\pm}$ 
&\multicolumn{2}{c}{ $\si_{\text{IBA}}^{pp/gg}$}
&\multicolumn{2}{c}{ $\si_{\text{IBA}}^{pp/q\bar q}$}
&\multicolumn{2}{c}{$\si_{\text{IBA}}^{pp/b\bar b}$}
&\multicolumn{2}{c}{$\si_{\text{IBA}}^{pp/g\ga}$}
&\multicolumn{2}{c}{$\De_{\text{EW}}^{pp/gg}$}
&\multicolumn{2}{c}{all}\\
\hline \hline
5  & 200 & 5&3652(9)  & 0&9885  &  0&0311  & 0&1058   & -0&2049(6) & 6&286(1)  \\ 
10 & 200 & 3&5138(6)  & 0&6551  &  0&0205  & 0&0695   & -0&3552(5) & 3&9037(8) \\
20 & 200 & 6&085(1)   & 1&095   &  0&035   & 0&119    &-1&367(2)   & 5&967(2)\\
10 & 300 & 1&2570(2)  & 0&0974  &  0&0058  & 0&0224   & -0&1292(2) & 1&2534(3)\\
10 & 400 & 0&5164(1)  & 0&0242  &  0&0019  & 0&0089   &-0&0544(1)  & 0&4971(1) \\
10 & 600 & 0&10583(2) & 0&00268 &  0&00027 & 0&00191  & -0&01295(2)& 0&09774(3) \\ 
\hline 
\end{tabular}\ec
 \end{footnotesize}
\end{table} 

\tab{table_ppHtb1} shows separately the  IBA results for the various
subprocesses together with the  
NLO EW corrections to $\ggHtbm$ at $\sqrt{s}= 14\tev$ for different values of $\MHpm$ and 
$\tan\beta$. Similar results are presented in \tab{table_ppHtb2} and
\tab{table_ppHtb3},
 but now for $\sqrt{s}= 8\tev$ and $\sqrt{s}= 7\tev$, respectively. 
We observe that the $gg$ contributions are dominant; 
they contribute more than $90\%$ ($83\%$) of the total IBA for 
$\sqrt{s}= 14\tev$  ($\sqrt{s}= 7\tev$). 
The contribution of the $b\bar b$ channel is below $1\%$;
the $g\ga$ channel contribution is slightly  larger.
The NLO EW contributions are comparable in size 
to the $q\bar q$ contributions, but with the opposite sign. 

\begin{figure}[t]
 \begin{center}
\includegraphics[width=7cm]{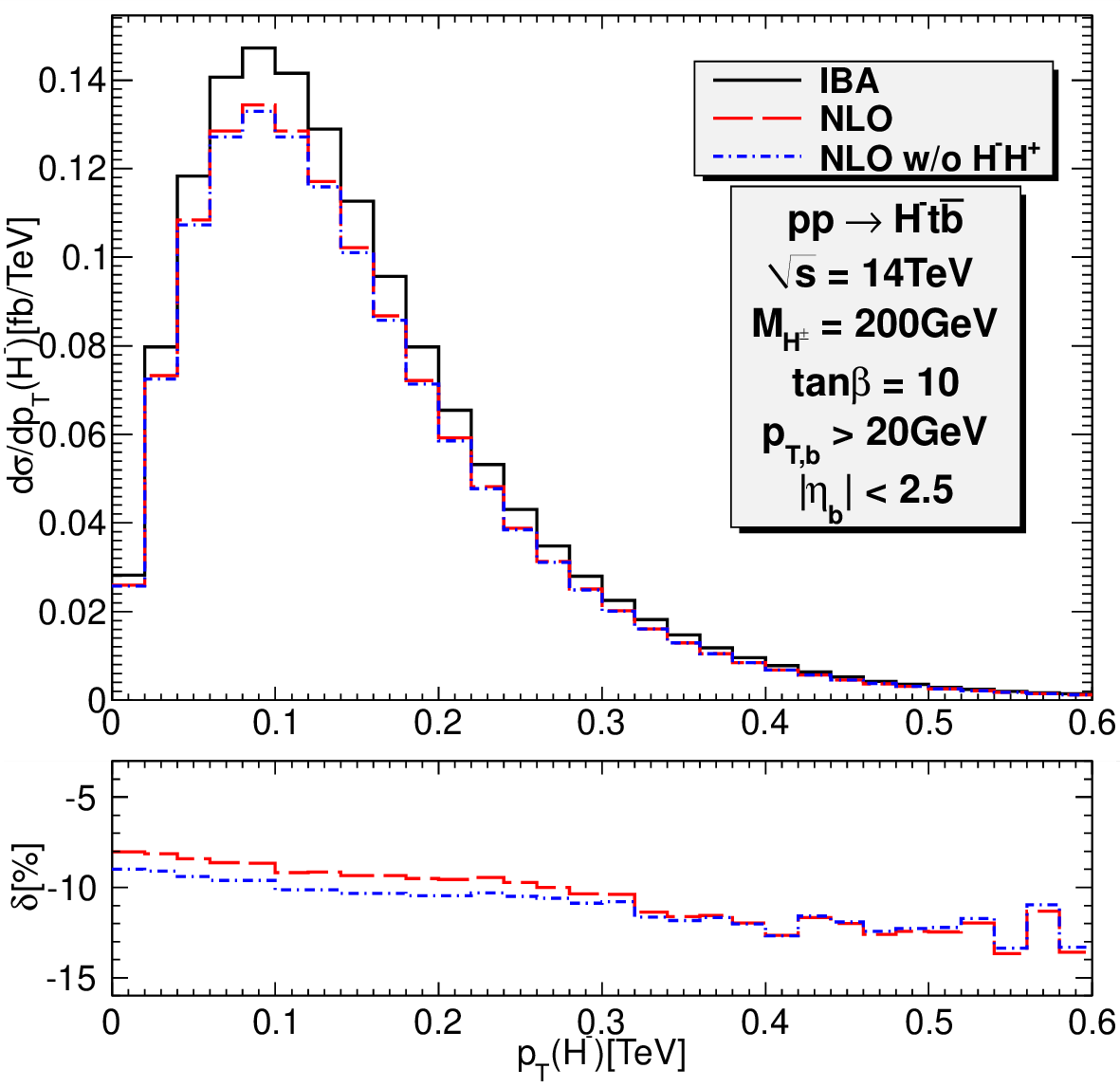}
\includegraphics[width=7cm]{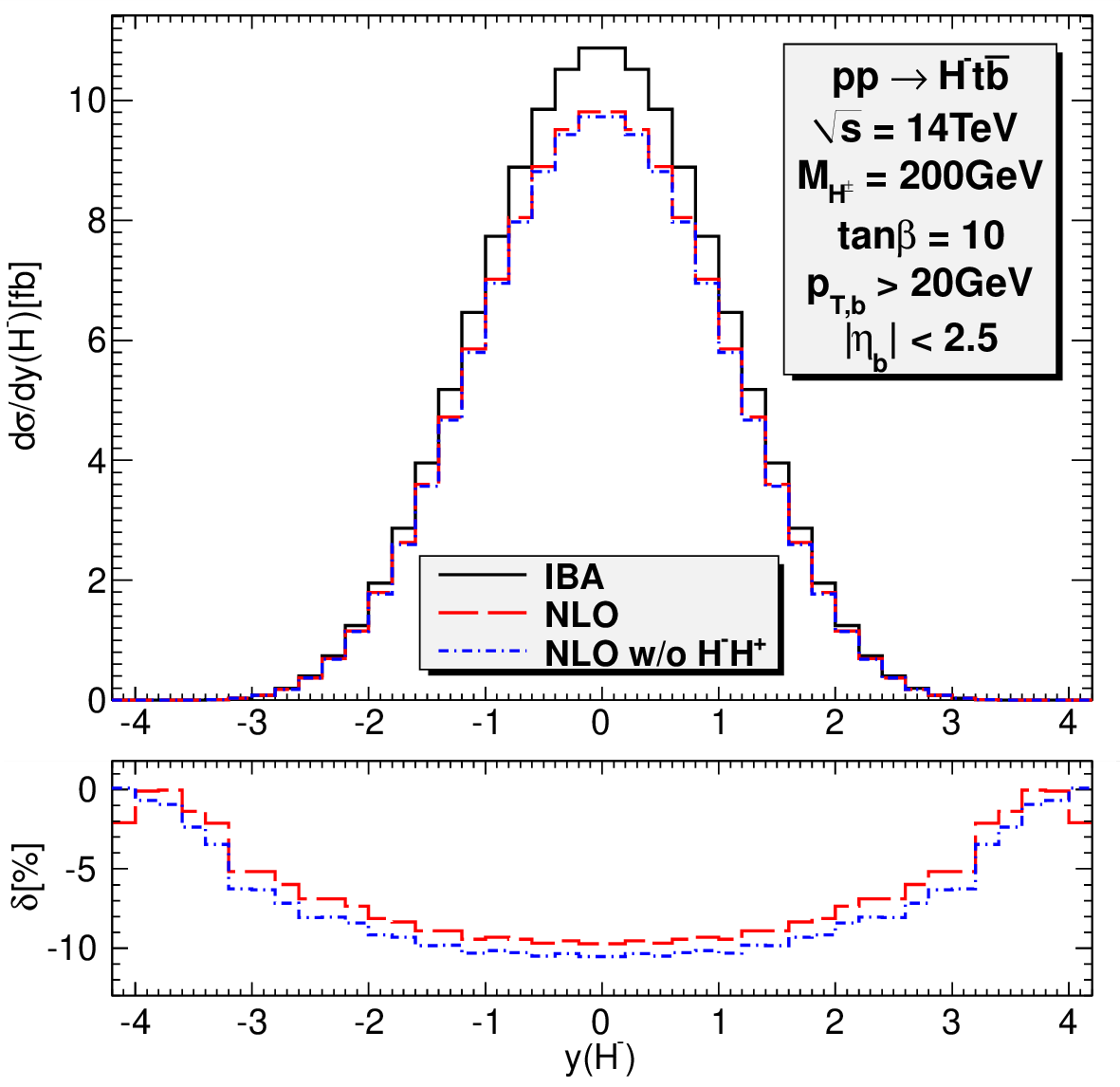}
\includegraphics[width=7cm]{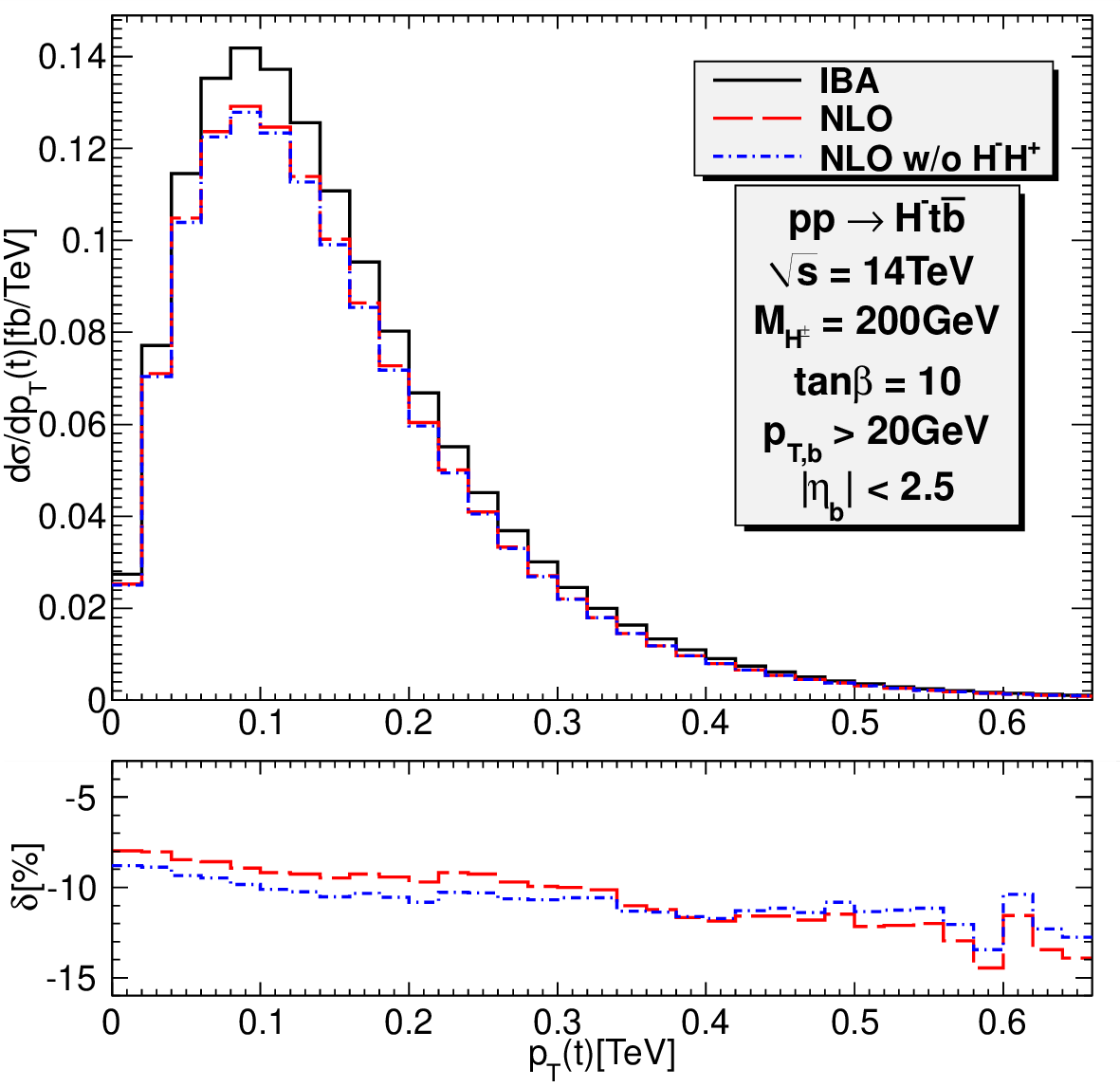}
\includegraphics[width=7cm]{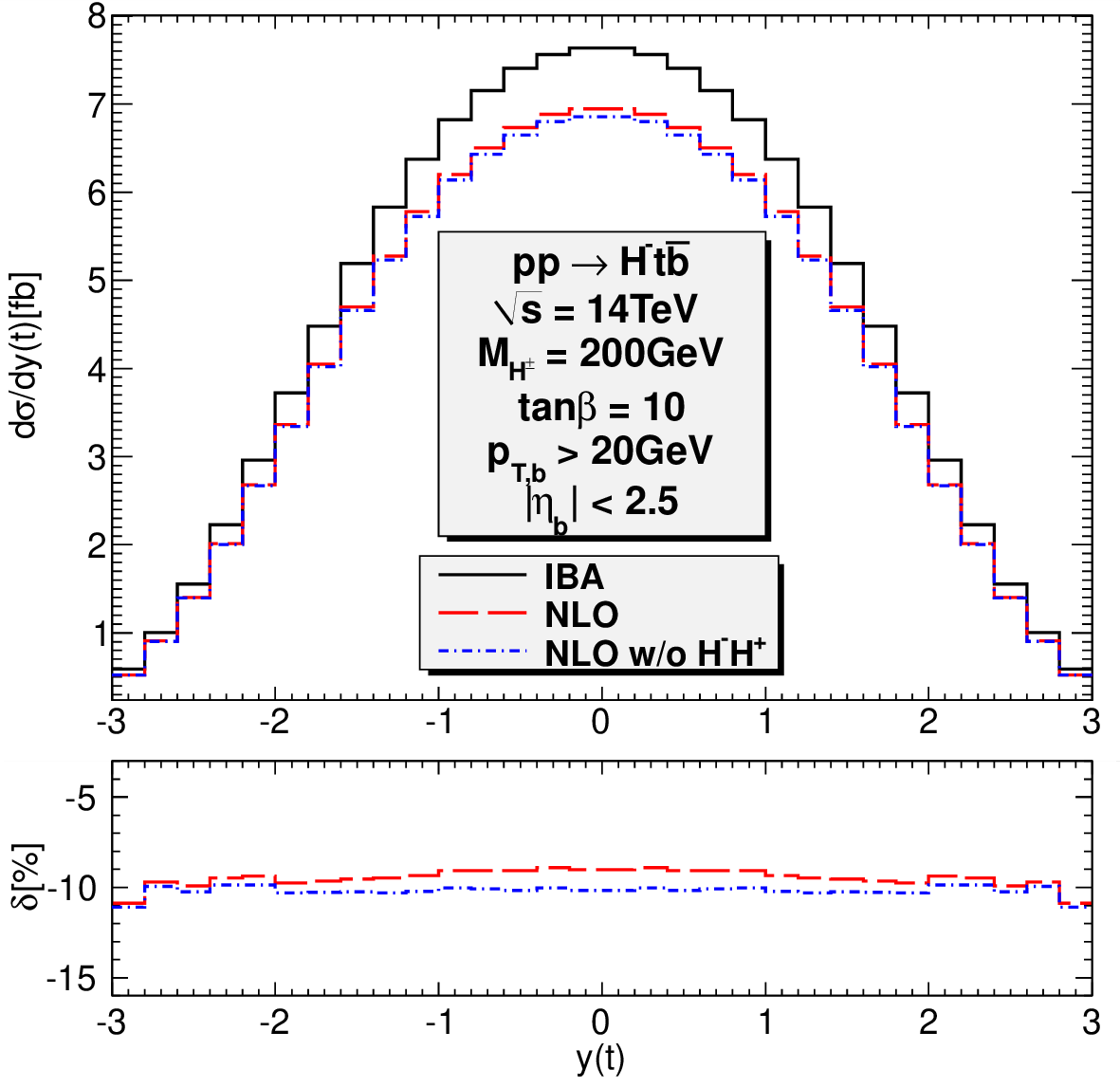}
\includegraphics[width=7cm]{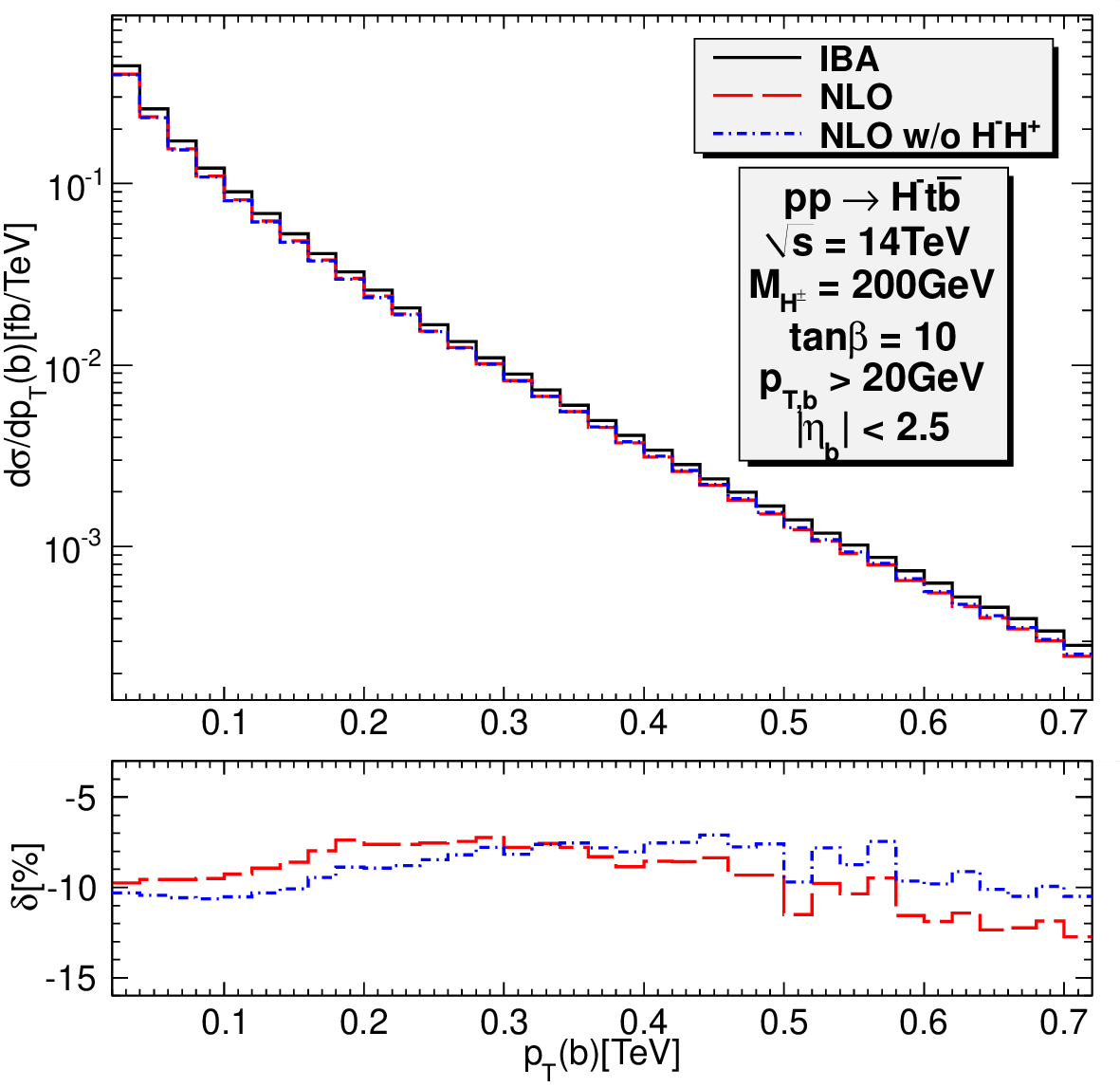}
\includegraphics[width=7cm]{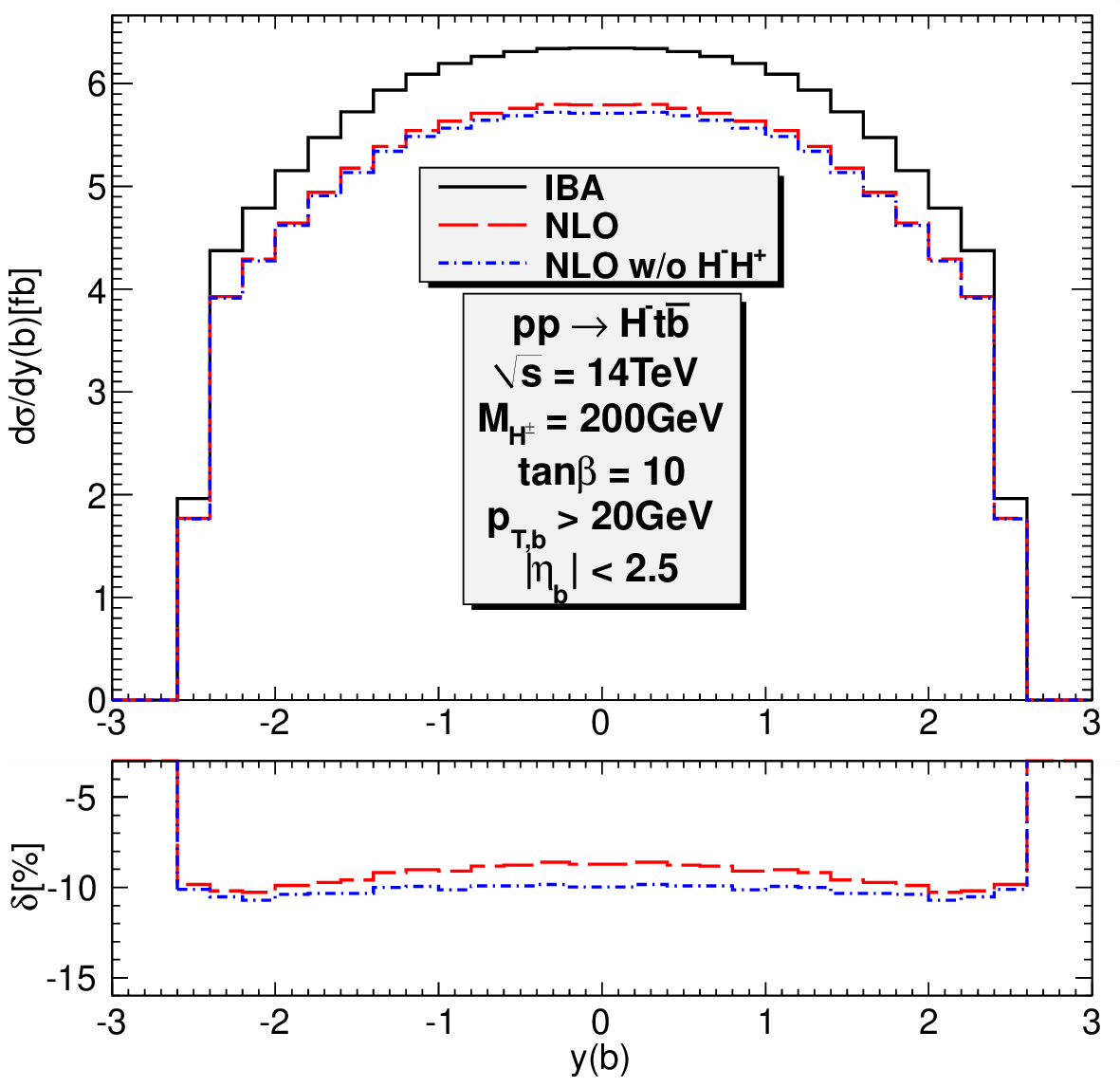}
\caption{
The IBA and NLO distributions of transverse momentum (left) and rapidity (right)
for $H^-$, $t$,  $\bar{b}$ at $14\tev$. 
The lower panels show the relative corrections. 
Also shown is the NLO result without the $H^-H^{+*}$ production mechanism.}
\label{fig:dist_Htb_14b}
\end{center}
\end{figure}

\begin{figure}[t]
 \begin{center}
\includegraphics[width=7cm]{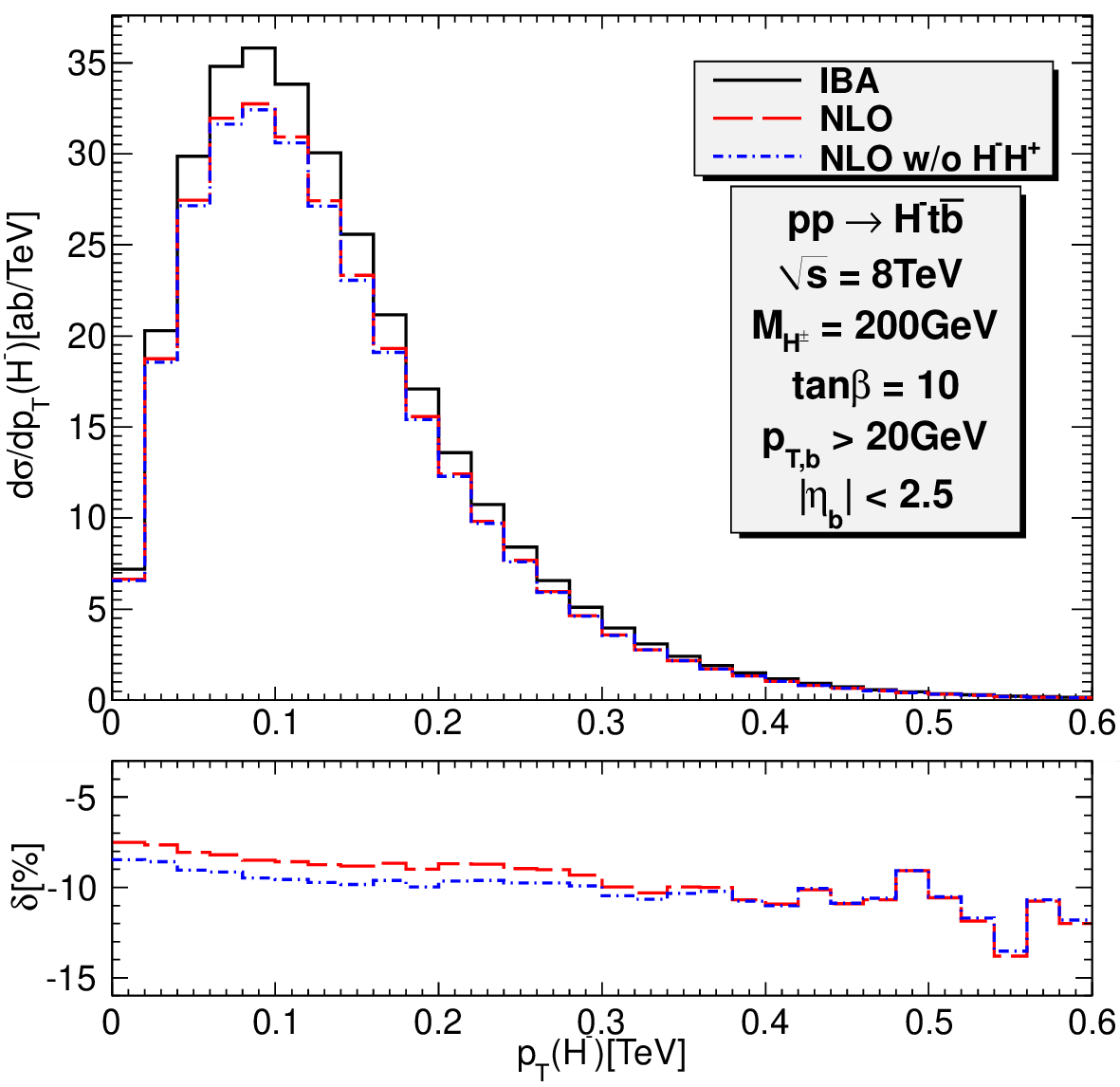}
\includegraphics[width=7cm]{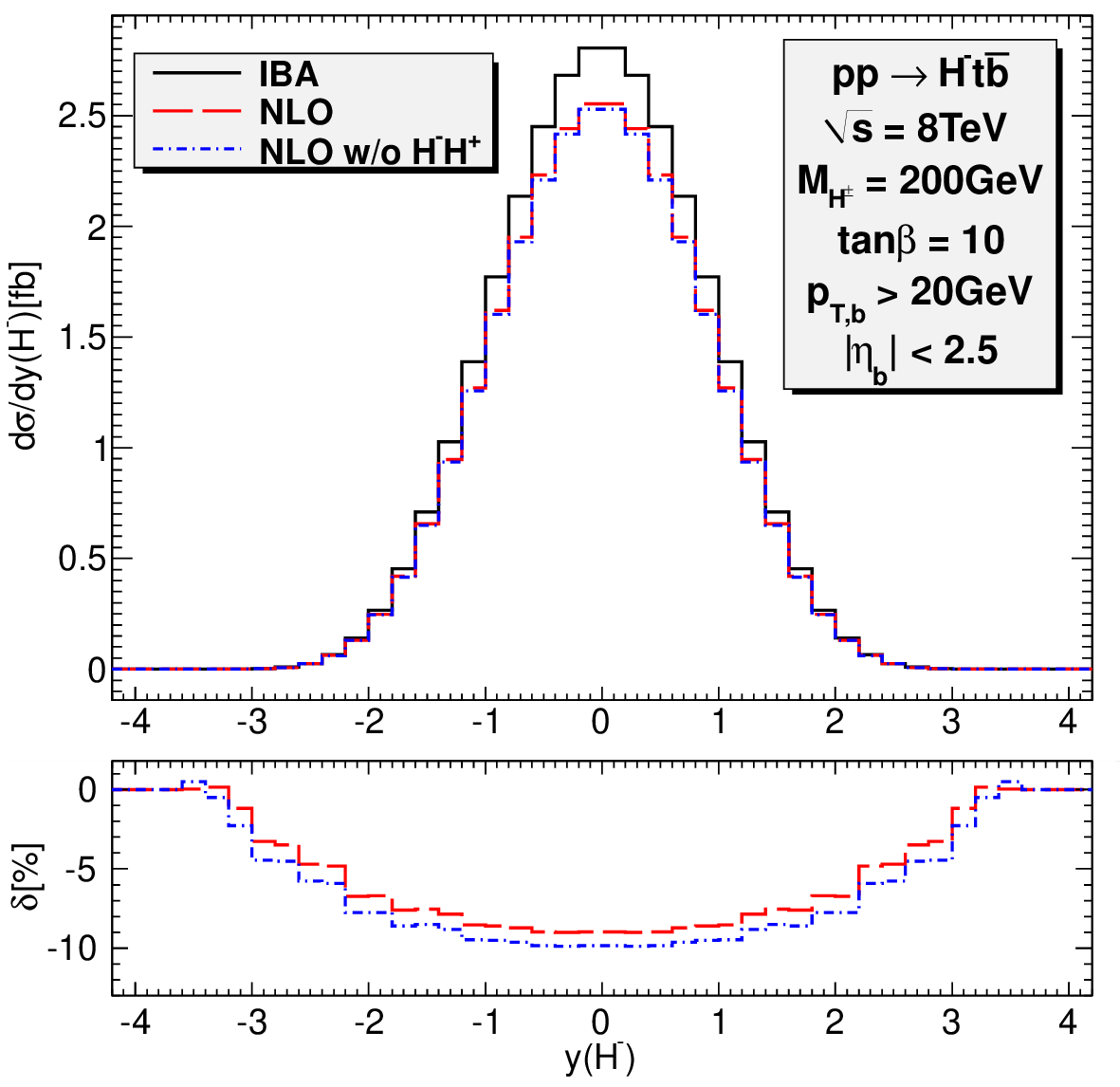}
\includegraphics[width=7cm]{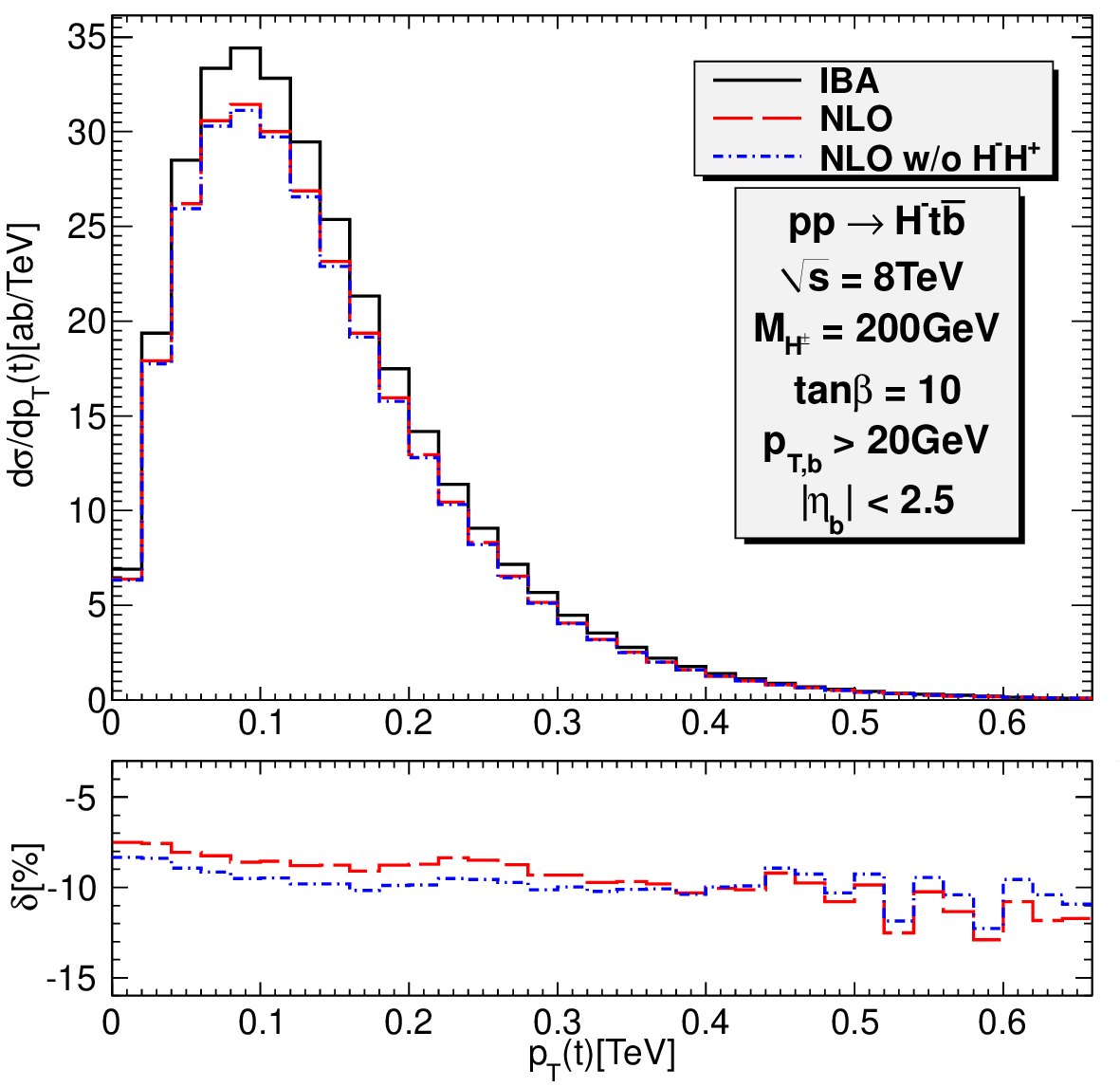}
\includegraphics[width=7cm]{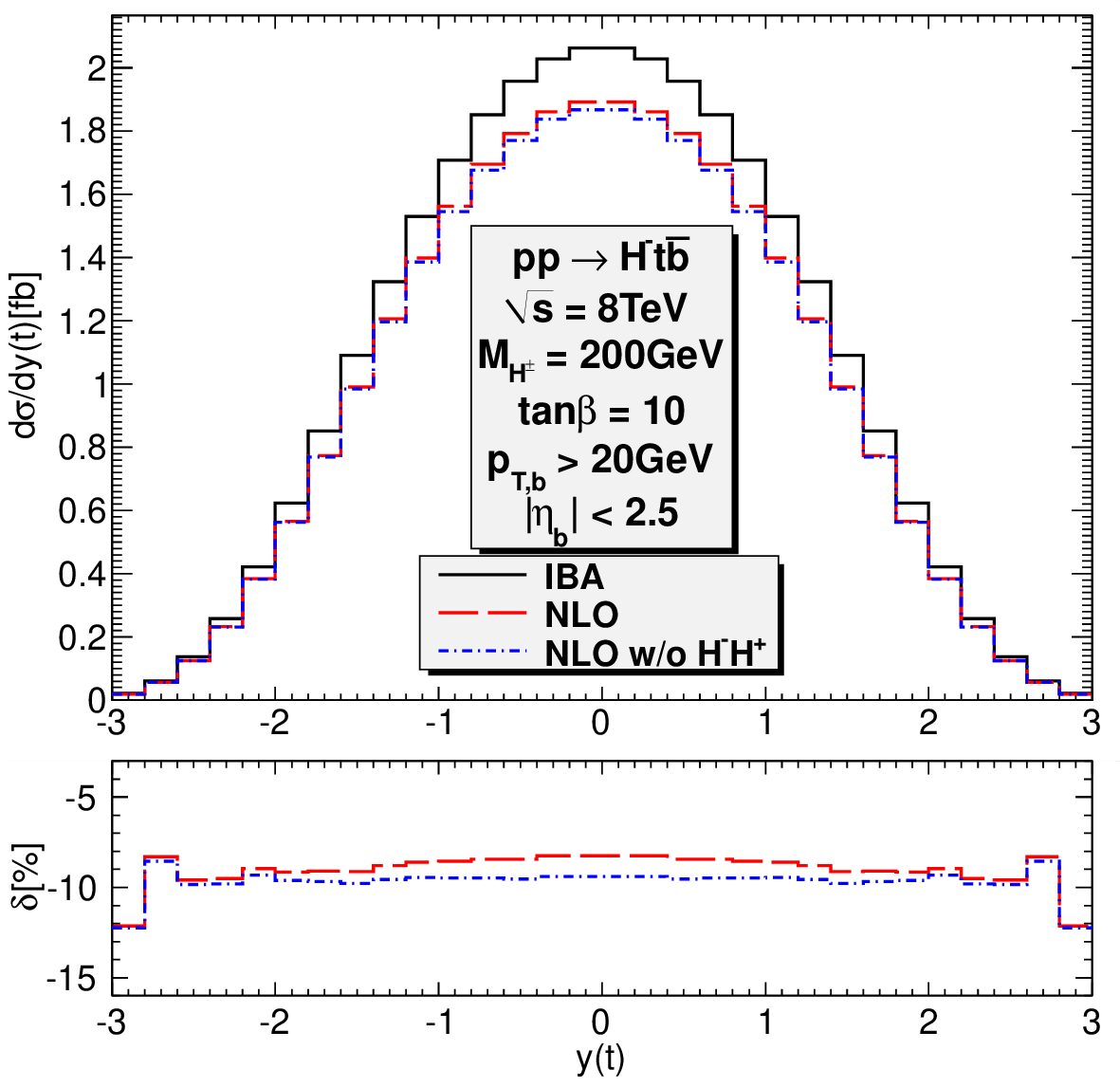}
\includegraphics[width=7cm]{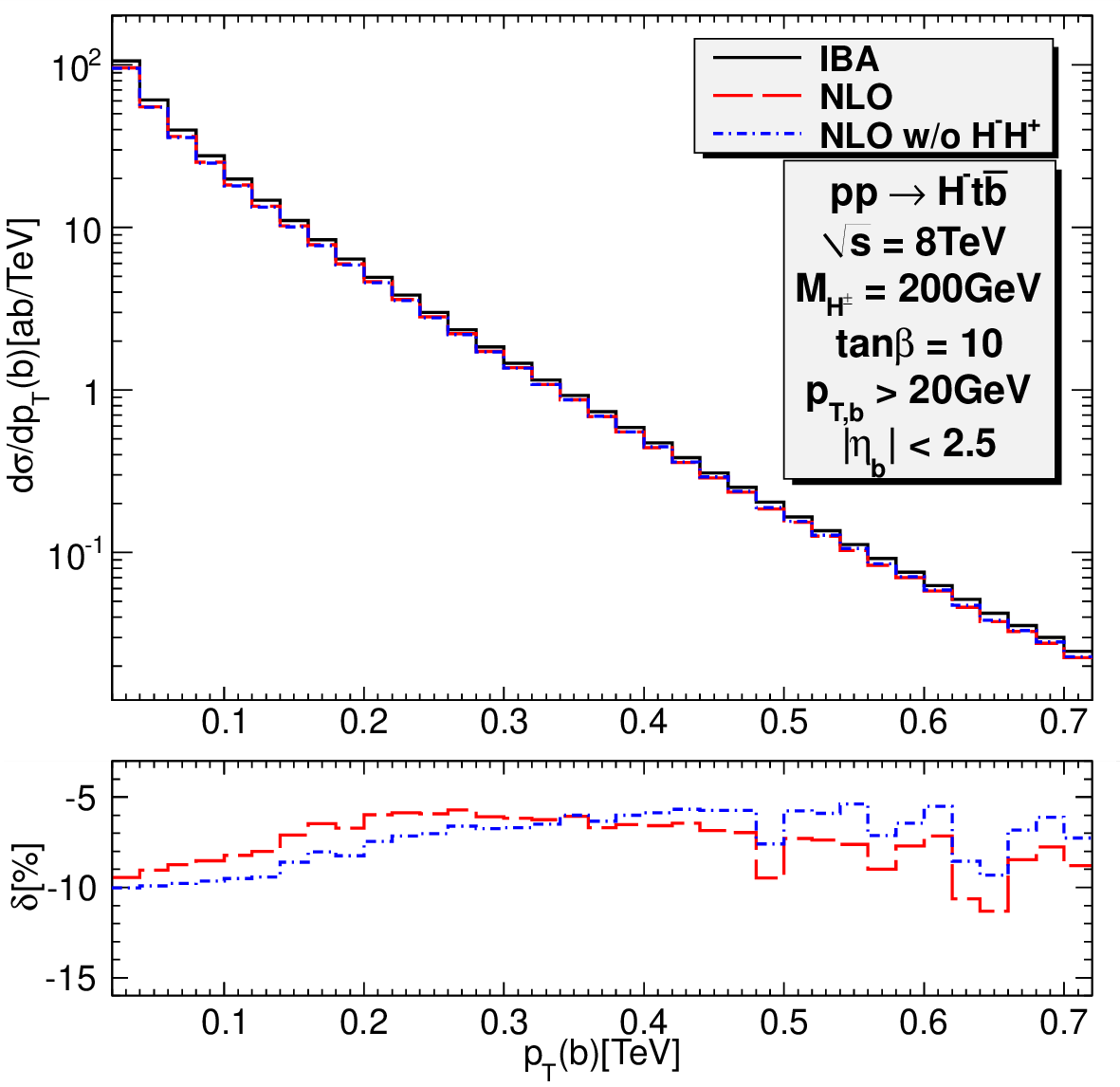}
\includegraphics[width=7cm]{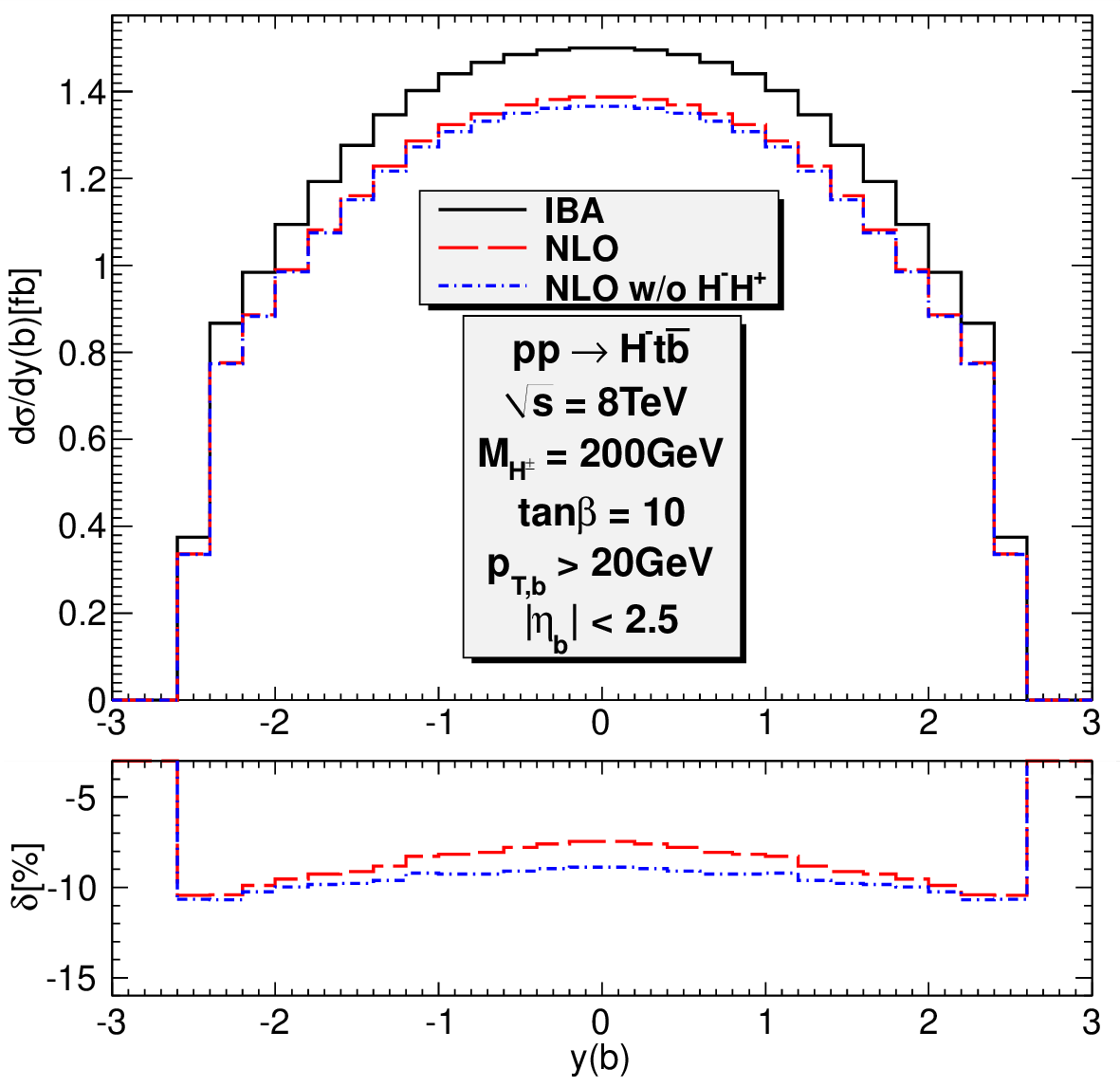}
\caption{Similar to \fig{fig:dist_Htb_14b} but for $\sqrt{s}=8\tev$.}\label{fig:dist_Htb_8b}
\end{center}
\end{figure}

\begin{figure}[t]
 \begin{center}
\includegraphics[width=7cm]{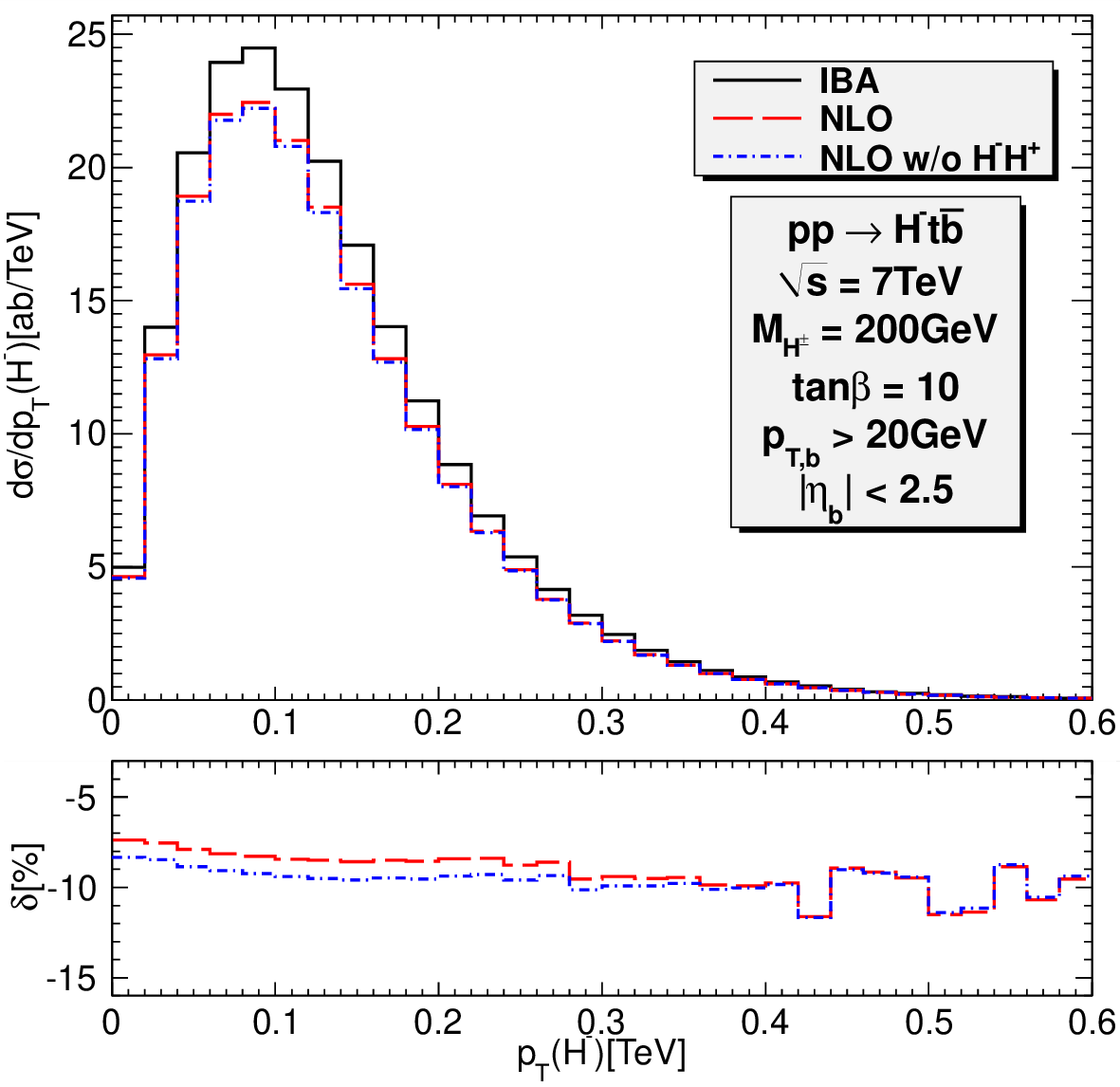}
\includegraphics[width=7cm]{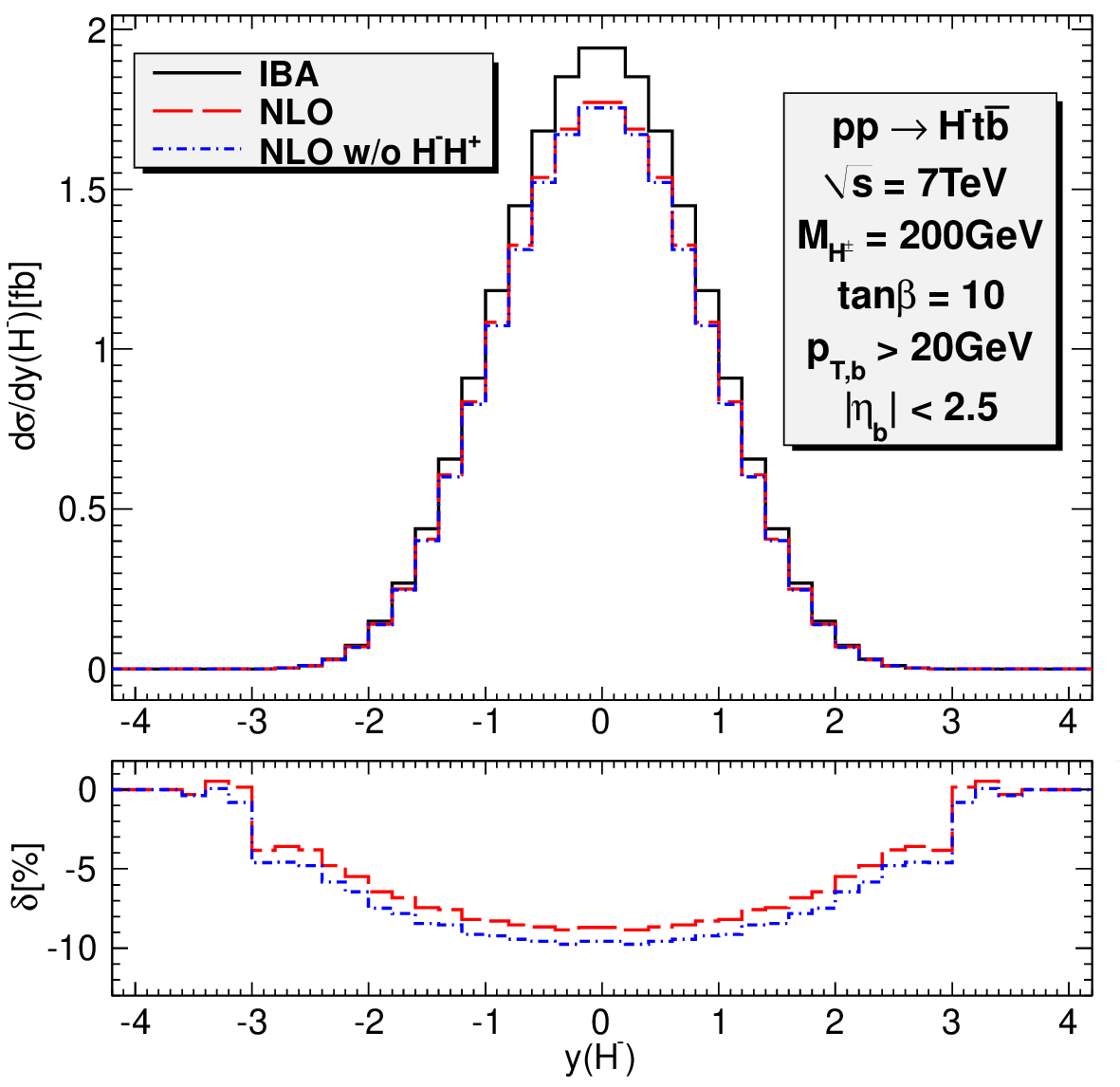}
\includegraphics[width=7cm]{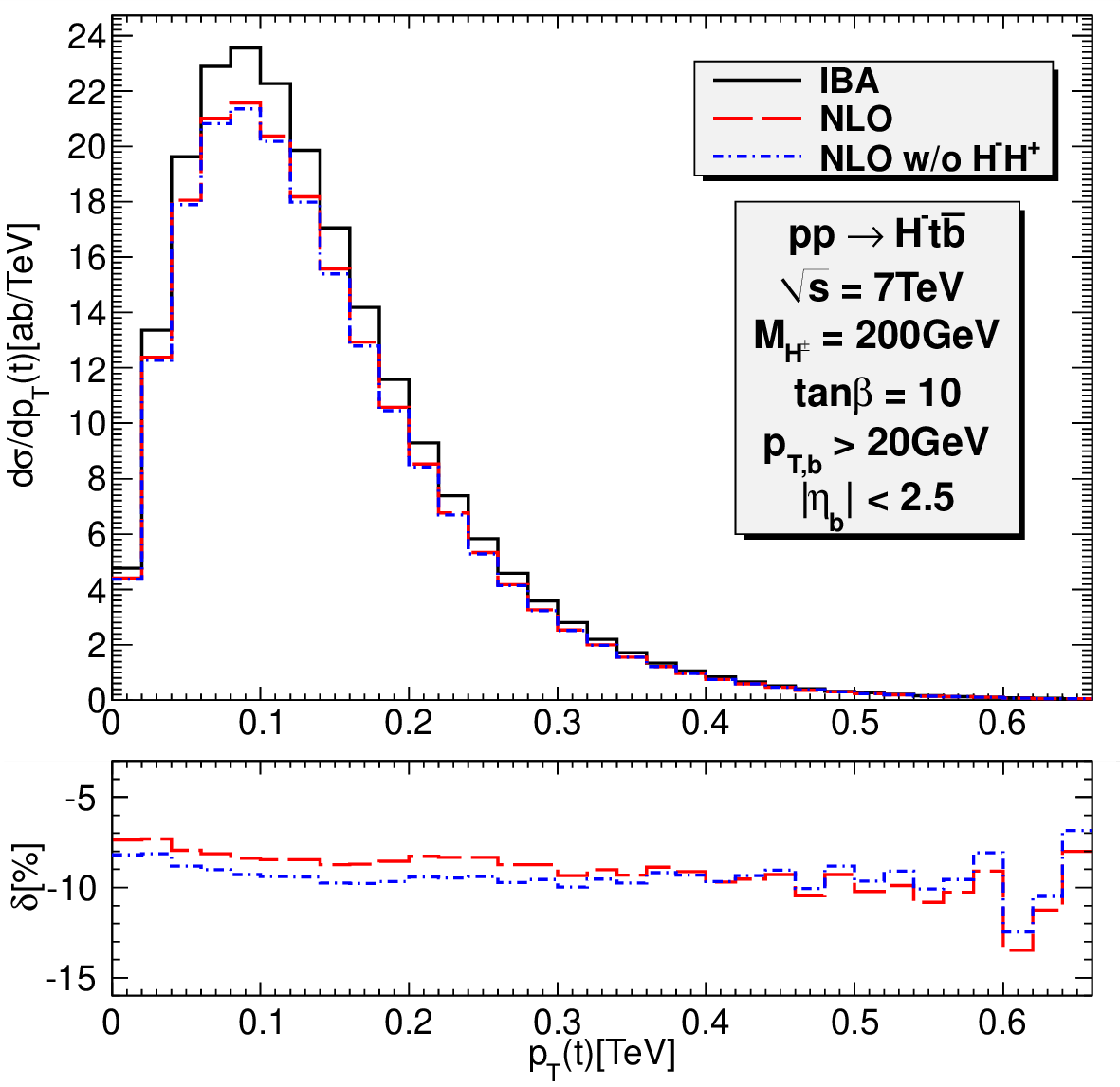}
\includegraphics[width=7cm]{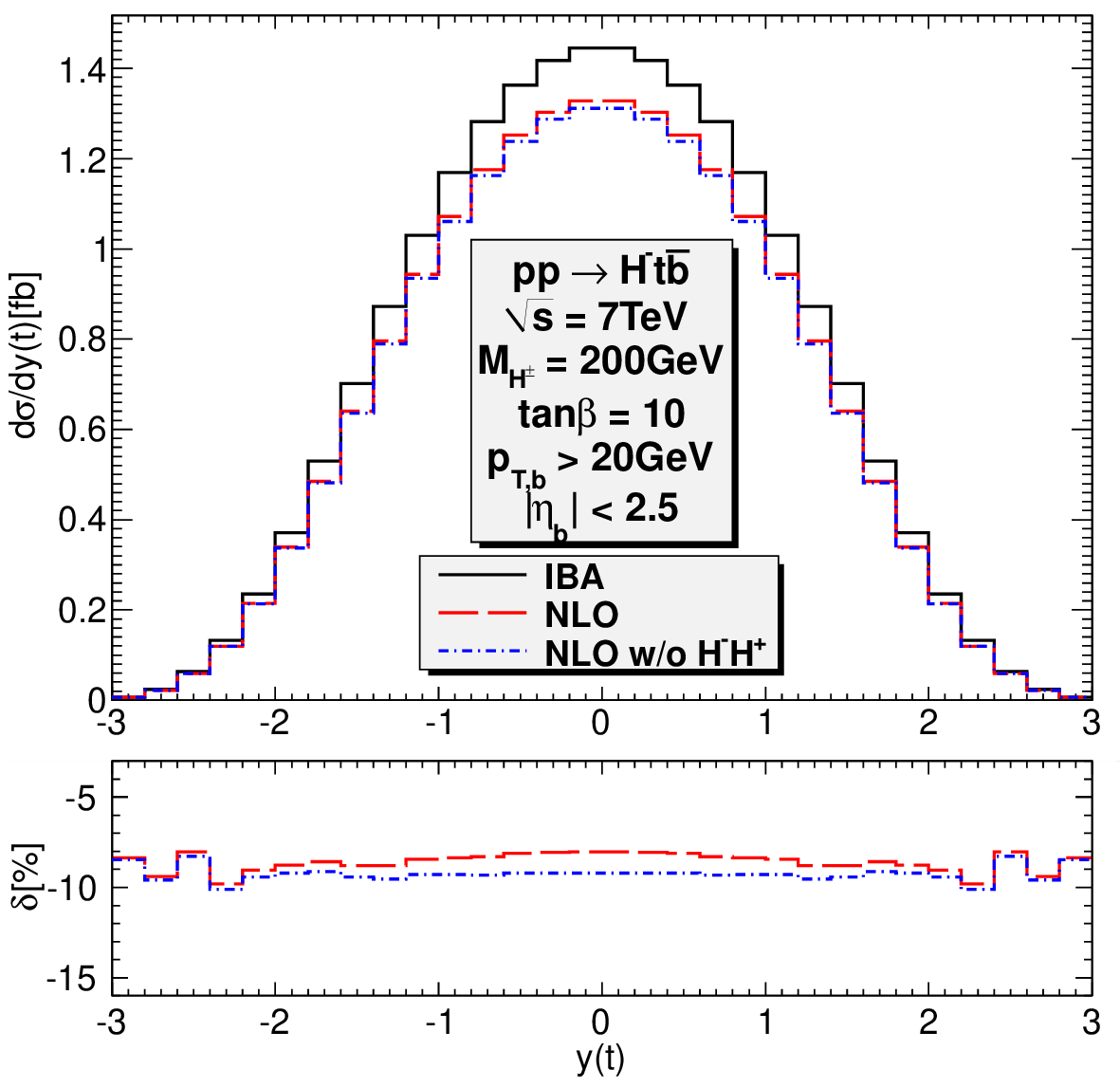}
\includegraphics[width=7cm]{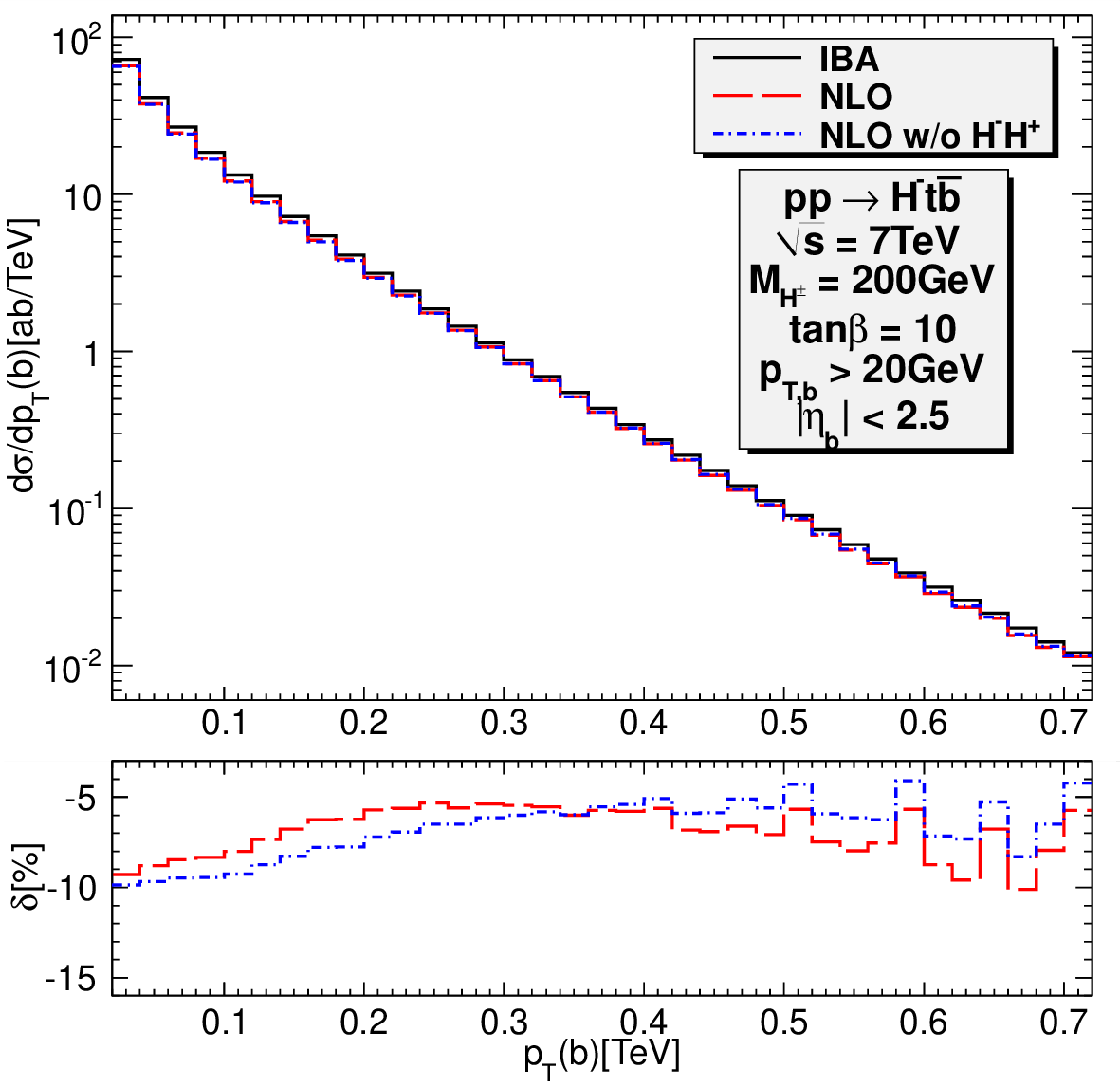}
\includegraphics[width=7cm]{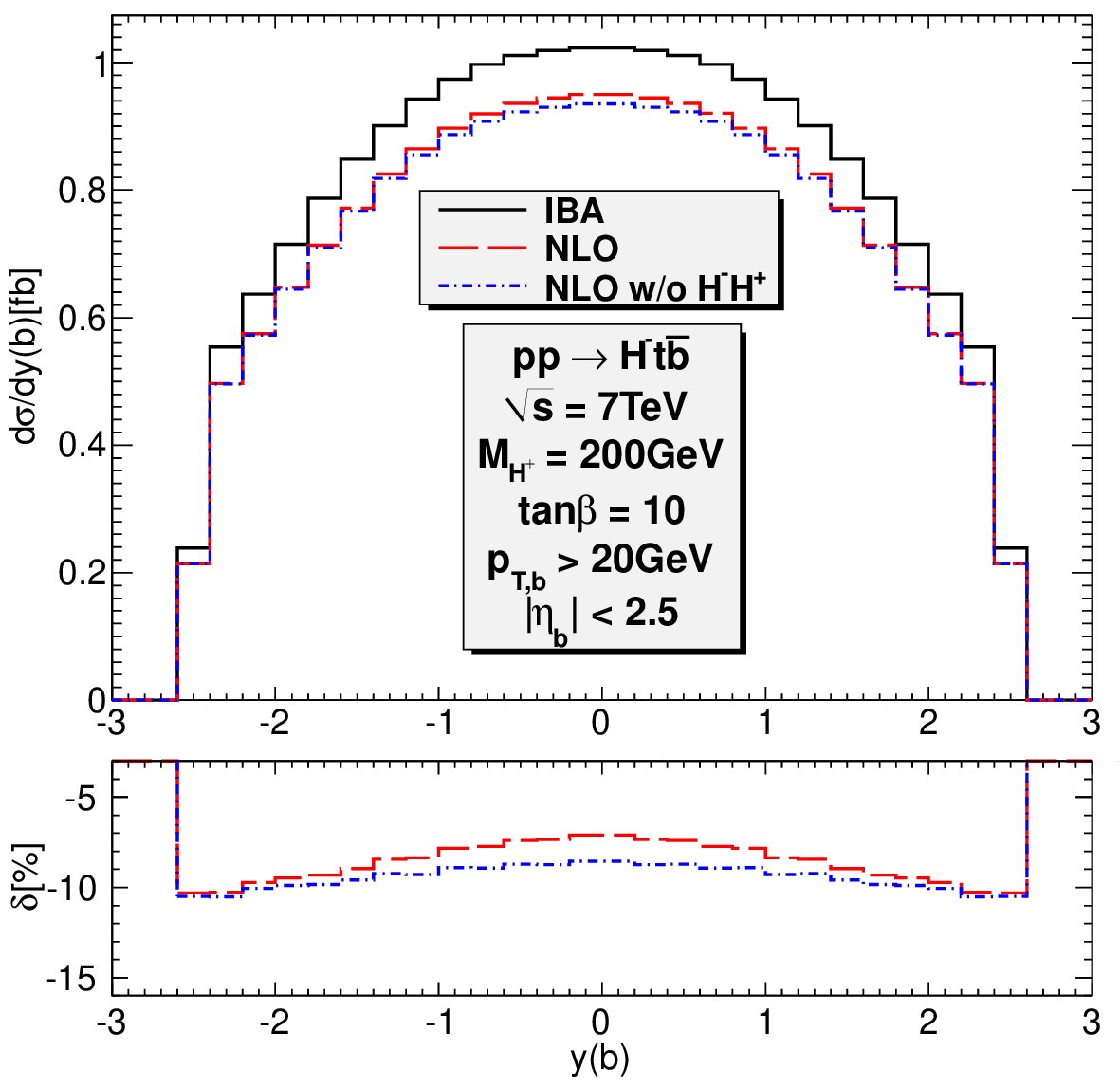}
\caption{Similar to \fig{fig:dist_Htb_14b} but for $\sqrt{s}=7\tev$.}\label{fig:dist_Htb_7b}
\end{center}
\end{figure}

\subsection{Differential distributions}
\label{sec:distributions}
We now consider the differential distributions of various kinematical  variables, 
in the  IBA and including the NLO EW corrections.
The relative correction is defined with respect to the IBA
differential cross section, 
$\delta = (d\sigma_{\text{NLO}} - d\sigma_{\text{IBA}})/d\sigma_{\text{IBA}}$. 
All results are shown in the right column of \fig{fig:dist_m45} 
and in Figs.~\ref{fig:dist_Htb_14b}, \ref{fig:dist_Htb_8b} and \ref{fig:dist_Htb_7b}. 

The effect of the $H^- H^{+*}$ production mechanism is best seen in the right column of 
\fig{fig:dist_m45}. 
The $t\bar{b}$ invariant mass distribution shows the singular pole 
structure at $M_{t\bar{b}}=M_{H^\pm}=200\gev$ if this channel is included. 
This effect is also visible in other distributions.

Distributions for the individual particles separately
are shown in \fig{fig:dist_Htb_14b}.
for $\sqrt{s}=14\tev$, and in Figs.~\ref{fig:dist_Htb_8b} and \ref{fig:dist_Htb_7b}
for the lower energies 8 and 7 TeV.
The results are very similar and differ essentially in the absolute
size of the cross section at the different energies.

For the charged Higgs boson, 
the relative correction is negative, decreases with $p_T(H^-)$ and 
has a minimum (about $-10\%$) at the central rapidity. 
 
For the top quark, the behavior 
 of the $p_T$ distribution is similar to the one of the charged Higgs boson. 
The EW corrections are negative and decrease with $p_T$, 
consistent with Sudakov corrections $\alpha\log^n(p_T^2/M_W^2)$ 
with $n=1,2$. For the rapidity  distribution, 
the relative correction is rather flat (about $-10\%$) in the region $|y_{t}|<3$. 
 
The distributions of the bottom quark are quite different from the ones of the heavy particles. 
At tree level (see the IBA curve), the cross section is larger at low $p_T$ due to collinear bottom-quark 
radiation off gluons. 
The relative correction increases and reaches the maximal value at $p_T\approx 0.3\tev$ and then follows 
the trend of decreasing with $p_T$ as for the other particles. This behavior can be explained by the 
interplay between the leading weak Sudakov correction
$\alpha\log^2(p_T^2/M_W^2)$ and the QED quasi-collinear 
correction $\alpha\log(m_b^2/p_T^2)$ from photon radiation off the bottom quark. 
The latter is more important at low $p_T$ while the former dominates 
in the high energy regime. 
For the rapidity distribution, the 
correction is smallest in  the central region.

\section{Conclusions}
\label{sect-conclusions}
In this paper we have studied the production of charged Higgs bosons in association
with a top quark and a tagged bottom quark at the LHC in the context of the complex MSSM. 
Cuts on the transverse momentum and rapidity of the bottom quark are applied. 
At tree level, the $gg$ fusion is dominant among various subprocesses
with quarks or photon in the initial state; for this parton process, 
the NLO EW corrections have been calculated and discussed.

Since the tree-level amplitudes are proportional to the top-bottom-Higgs coupling, 
we have examined the effective-coupling approximation and compared it to 
the full NLO result. 
The dependence of the cross section on $\tan\beta$, $M_{H^\pm}$ and 
the phase $\phi_t$ of the trilinear coupling $A_t$ has also been studied. 

Numerical results have been presented for the CPX scenario. The
production cross section shows a strong dependence on $\tan\beta$,
$M_{H^\pm}$ and $\phi_t$. Large production rates occur
for small $\tan\beta$, small $M_{H^\pm}$ and phases $\phi_t$
around $\pm \pi$. At LO, the cross section increases strongly with large $\tan\beta$. 
This behavior is, however, significantly reduced when NLO corrections are included. 
An interesting feature is the $\phi_t$ dependence: while the LO cross section is just 
a constant, the IBA and NLO results show a strong dependence with a minimum at $\phi_t=0$. 

We have also presented various differential distributions of the final state particles,
where the NLO EW corrections are usually negative. 

\vspace{0.7cm}
\noindent {\bf Acknowledgments} \\
D.T.N. and L.D.N. would like to thank the Max-Planck Insitut f{\"u}r Physik in Munich where 
most of this work has been done and acknowledge the support from the Deutsche Forschungsgemeinschaft
via the Sonderforschungsbereich/Transregio SFB/TR-9 Computational Particle Physics.
L.D.N. is partially supported by the Vietnam Academy of Science and Technology between the Vietnam-France collaboration program in particle physics under the grant VAST.HTQT.PHAP.04/2012-2013.



\end{document}